\documentclass[onecolumn,authoryear]{els-mrw} 

\usepackage{amsmath,amssymb,amsfonts,amsthm,makeidx,graphicx}
\usepackage{bm}
\usepackage{txfonts}
\usepackage{helvet}
\usepackage{hyperref}[colorlinks=blue, citecolor=blue, linkcolor=blue, urlcolor=blue]
\usepackage{color}

\begin{document}

\chapter{Higher Twists in QCD}\label{chap1}

\author[1]{W. Melnitchouk}
\address[1]{\orgname{Theory Center}, \orgdiv{Jefferson Lab}, \orgaddress{Newport News, Virginia 23606, USA}}

\maketitle

\begin{glossary}[Nomenclature]
\begin{tabular}{@{}lp{34pc}@{}}
CFF & Compton form factor           \\
DGLAP & Dokshitzer-Gribov-Lipatov-Altarelli-Parisi \\
DIS & Deep-inelastic scattering     \\
DVCS & Deeply-virtual Compton scattering \\
GPD & Generalized parton distribution \\
JAM & Jefferson Lab Angular Momentum (Collaboration)\\
OPE & Operator product expansion    \\
PDF & Parton distribution function  \\
QCD & Quantum chromodynamics        \\
SIDIS & Semi-inclusive deep-inelastic scattering     \\
TMC & Target mass correction        \\
TMD & Transverse momentum dependent
\end{tabular}
\end{glossary}

\begin{abstract}[Abstract]
Higher twists in deep-inelastic scattering (DIS) and related hard scattering processes provide a unique window on nonperturbative QCD dynamics, encoding quark--gluon correlations and multiparton interactions beyond the leading-twist parton model. We present an overview of the theoretical foundations of higher twists in QCD, based on the operator product expansion, and discuss their intimate connection with quark--hadron duality in the transition between the resonance and scaling regimes. We review the phenomenology of higher-twist effects in unpolarized and polarized DIS, emphasizing their extraction from precision data through modern global QCD analyses that simultaneously determine both leading-twist parton distribution functions and higher-twist contributions. Finally, we discuss recent developments in the study of higher twists in semi-inclusive and exclusive reactions, and conclude by highlighting future opportunities to study QCD dynamics beyond leading twist at Jefferson Lab and the Electron--Ion Collider that promise to deepen our understanding of hadron structure.
\end{abstract}


\begin{BoxTypeA}{Key points}
\begin{itemize}
\item The operator product expansion provides a systematic framework for describing power-suppressed multiparton correlations in QCD.
\item Structure function moments reveal a fundamental connection between higher-twist dynamics and quark–hadron duality.
\item Precision measurements and global QCD analyses can constrain multiparton correlations in unpolarized and polarized deep-inelastic scattering.
\item Semi-inclusive and exclusive reactions offer complementary probes of quark–gluon dynamics beyond the leading-twist approximation.
\end{itemize}
\end{BoxTypeA}

\section{Introduction}
\label{sec:intro}

Deep-inelastic scattering (DIS) has played a foundational role in the development of our understanding of QCD and the internal structure of hadrons. Beginning with the pioneering electron--proton scattering experiments at SLAC in the late 1960s, measurements of inclusive structure functions revealed the predicted approximate scaling [\cite{Bjorken1969}] and provided compelling evidence for pointlike constituents inside the nucleon [\cite{Bloom1969, Breidenbach1969}]. These observations led naturally to the parton model~[\cite{Feynman1969, Feynman1972}], in which the exchanged virtual photons scatter incoherently from nearly free quarks and gluons (partons) inside the nucleon. With the advent of QCD~[\cite{GrossWilczek1973, Politzer1973}], the approximate scaling observed in DIS was understood as a consequence of asymptotic freedom, whereby the strong coupling becomes small at large momentum transfers, rendering quarks effectively free during the hard scattering. The subsequent observation of logarithmic scaling violations~[\cite{GeorgiPolitzer1974, GrossWilczek1974}] was instrumental in establishing QCD as the theory of the strong interaction and provided the theoretical framework for describing high-energy lepton--hadron scattering.

Within perturbative QCD, inclusive DIS structure functions are expressed in terms of universal parton distribution functions (PDFs) convoluted with perturbatively calculable hard-scattering kernels~[\cite{Collins:1989gx, Collins:2011zzd}]. The scale dependence of PDFs is governed by the DGLAP evolution equations~[\cite{GribovLipatov1972, AltarelliParisi1977, Dokshitzer1977}], allowing measurements over a wide range of momentum transfers to be related through the renormalization group. Over the past several decades, this framework has achieved remarkable success in describing a vast body of experimental data from accelerator facilities worldwide. Modern global QCD analyses~[\cite{CT18, MSHT20, NNPDF40, Cocuzza:2026zoy, Alekhin:2017kpj}] determine PDFs with increasingly high precision, providing critical input for virtually every quantitative prediction involving hadronic processes. At sufficiently large momentum transfers, the leading-twist approximation, supplemented by perturbative QCD evolution, provides an excellent description of experimental measurements.

The increasing precision of both experiments and theoretical calculations, however, has shifted attention toward effects that were once regarded primarily as corrections to, or even obscuring, the underlying leading-twist dynamics. Modern measurements extend well beyond the asymptotic Bjorken scaling regime into regions of moderate momentum transfer squared, $Q^2 \sim 1$~GeV$^2$, and large values of the Bjorken scaling variable $x_B$, where contributions that are nominally suppressed by inverse powers of the hard scale can become numerically significant. In this kinematic regime the simple leading-twist description is no longer sufficient, and a quantitative understanding requires the inclusion of target-mass corrections, threshold effects, and genuine higher-twist contributions arising from multiparton correlations. Rather than representing unwanted corrections to be removed from the data, these power-suppressed terms provide unique information on the nonperturbative dynamics of QCD and the correlations among quarks and gluons inside nucleon.

The usual theoretical framework for organizing such corrections is provided by the operator product expansion (OPE)~[\cite{Wilson:1969zs}]. In the OPE, moments of structure functions are expanded systematically in powers of $1/Q^2$, with the coefficients of each term given by matrix elements of local operators of definite twist, defined as the difference between the canonical dimension and spin of the operator. The leading-twist contribution describes incoherent scattering from individual partons, while higher-twist terms arise from operators involving correlated quark and gluon fields and therefore encode genuine multiparton correlations. The OPE therefore provides a direct connection between experimentally measurable moments of structure functions and the matrix elements of higher-twist operators. Moments also smooth the detailed $x_B$ dependence of structure functions and naturally average over resonance structure, making them especially useful for studying the transition between perturbative and nonperturbative QCD. Furthermore, low-order moments are increasingly accessible from lattice QCD calculations, providing one of the most direct quantitative connections between experimental measurements and first-principles calculations in QCD.

The intimate relationship between higher twists and moments leads naturally to the phenomenon of quark--hadron duality, which has been the subject of extensive theoretical and experimental investigation over the past five decades~[\cite{Melnitchouk:2005zr}]. First observed in the early SLAC DIS experiments in the late 1960s~[\cite{Bloom:1970xb, Bloom:1971ye}], duality refers to the remarkable empirical observation that suitably averaged resonance-region structure functions closely follow the scaling curves measured in the deep-inelastic continuum. Although this phenomenon was discovered before the formulation of QCD, it was subsequently understood to emerge naturally within the framework of the OPE, in which moments of structure functions are organized according to operators of increasing twist~[\cite{DeRujula:1976baf}]. Importantly, quark--hadron duality does not imply the absence of higher-twist effects. Rather, it reflects the fact that the net contribution of higher-twist terms to suitably averaged observables is often substantially reduced, even though individual resonances may exhibit significant nonperturbative dynamics~[\cite{DeRujula:1976baf, Ji:1994br, Melnitchouk:2005zr}]. The concepts of higher twists, moments, and quark--hadron duality are therefore not independent subjects, but complementary manifestations of the same underlying QCD dynamics.

The past two decades have witnessed renewed interest in higher twists, driven largely by the availability of increasingly precise measurements from Jefferson Lab~[\cite{Dudek2012}]. The 6~GeV and, more recently, 12~GeV experimental programs have produced extensive data at large $x_B$ and moderate $Q^2$, enabling detailed studies of resonance-region duality and moments of structure functions~[\cite{Niculescu:2000tj, Niculescu:2000tk}], and motivating global QCD analyses that extend well beyond the kinematic region traditionally used in PDF extractions by incorporating target mass, nuclear, and higher-twist corrections within a unified framework~[\cite{Accardi:2016qay, Cerutti:2025yji, Accardi:2026hdv, Cocuzza:2025qvf, Cocuzza:2026zoy}]. At the same time, the extraction of higher-twist contributions remains intrinsically challenging because apparent power corrections can arise from a variety of sources, including finite target masses, threshold effects, higher-order perturbative corrections, soft-gluon resummation, and limitations of leading-twist parametrizations. Understanding these correlations has become an increasingly important aspect of modern precision QCD phenomenology.

Higher-twist effects also play an important role beyond inclusive DIS. Power corrections contribute to semi-inclusive DIS (SIDIS), to transverse momentum dependent (TMD) observables through $qqg$ correlations, and to exclusive reactions through twist-three generalized parton distributions and meson distribution amplitudes. At the same time, rapid progress in lattice QCD has made possible increasingly precise calculations of moments of leading-twist distributions, twist-three matrix elements such as $d_2$, and exploratory studies of higher-twist operators. Together with future measurements at Jefferson Lab and the Electron--Ion Collider, these developments promise significant advances in our understanding of nonperturbative QCD over the coming decade.

We begin our discussion in Sec.~\ref{sec:DIS} by reviewing the formalism of inclusive DIS, QCD factorization, and the twist expansion of structure functions. We outline both the OPE framework and the diagrammatic approach, both of which provide the theoretical underpinning for a systematic description of power corrections in hard scattering processes in modern QCD phenomenology. In Sec.~\ref{sec:duality} we follow this by examining the phenomenon of quark--hadron duality, highlighting its intimate connection with moments and higher twists and its interpretation within the OPE. The phenomenology of higher twists is the subject of Sec.~\ref{sec:pheno}, where we discuss their physical origin, theoretical description, and determination in modern global QCD analyses of experimental data. Extensions beyond inclusive DIS are reviewed in Sec.~\ref{sec:beyond}, including higher-twist effects in SIDIS, TMD distributions and generalized parton distributions (GPDs). Finally, Sec.~\ref{sec:summary} summarizes the principal conclusions of this review and discusses future opportunities for advancing our understanding of higher-twist dynamics through increasingly precise global analyses of data from the Jefferson Lab 12~GeV program and the future Electron--Ion Collider.

\section{Deep-Inelastic Scattering and the Twist Expansion}
\label{sec:DIS}

Historically, inclusive DIS has provided the theoretical and experimental framework within which higher-twist effects have been most extensively studied. In this section we establish the formalism and notation that will be used throughout the remainder of this review. We begin with a summary of the kinematics of inclusive DIS and definitions of hadronic structure functions that parametrize the interaction of the virtual photon with the target hadron. We then briefly review the QCD factorization framework, which relates the leading-twist structure functions to universal PDFs and perturbatively calculable coefficient functions, and introduce the twist expansion, which organizes contributions to the structure functions according to their suppression by inverse powers of the hard scale. Particular emphasis is placed on distinguishing leading-twist contributions from kinematic power corrections arising from finite target masses and from genuine dynamical higher twists associated with quark--gluon and multi-quark correlations. These concepts provide the foundation for the discussion of the OPE, moments of structure functions, and the phenomenology of higher twists in the following sections.

\subsection{Inclusive DIS}

Inclusive deep-inelastic lepton--nucleon scattering remains the cornerstone of our knowledge of nucleon structure and the primary source of information on the PDFs that enter essentially all high-energy processes involving hadrons~[\cite{Close:1979, Roberts:1990, Thomas:2001}]. Because the electromagnetic interaction is perturbatively well understood, the measured cross sections in principle provide a relatively clean probe of the hadronic structure encoded in the nucleon structure functions.

In the one-photon exchange approximation, the inclusive scattering process,
\begin{equation}
\ell(k) + N(P) \to \ell'(k') + X(p_X),
\label{eq:DISprocess}
\end{equation}
is mediated by the exchange of a virtual photon (or weak boson) carrying four-momentum $q=k-k'$, where $k$ and $k'$ denote the four-momenta of the incoming ($\ell$) and outgoing ($\ell'$) leptons, respectively, and $P$ is the momentum of the target nucleon. The exchanged photon is spacelike, $Q^2=-q^2>0$, and the energy transfer in the target rest frame is $\nu=(P\cdot q)/M$, where $M$ is the nucleon mass. The invariant mass squared of the produced hadronic system is
\begin{equation}
W^2 \equiv p_X^2 = (P+q)^2 = M^2 + Q^2 \left( \frac{1}{x_B}-1 \right),
\label{eq:W2}
\end{equation}
where 
\begin{equation}
x_B = \frac{Q^2}{2P\cdot q}
\label{eq:xBj}
\end{equation}
is the Bjorken scaling variable, which lies in the interval $0 \le x_B \le 1$. In the parton model, or at leading order in perturbative QCD, $x_B$ can be identified with the fraction of the nucleon's longitudinal momentum, $x$, carried by the struck quark. The total squared center-of-mass energy squared, $s=(k+P)^2$, is related to the photon virtuality by
\begin{equation}
Q^2 = x_B\, y \left(s-M^2\right),
\end{equation}
where $y = (P\cdot q)/(P\cdot k)$ is the lepton inelasticity.

The double differential cross section for spin-averaged inclusive DIS can be written in the form
\begin{equation}
\frac{d^2\sigma}{dE'd\Omega}
= \frac{\alpha^2}{Q^4} \frac{E'}{E} L_{\mu\nu}W^{\mu\nu},
\label{eq:DISxsec}
\end{equation}
where $L_{\mu\nu}$ is the leptonic tensor, which is calculable from quantum electrodynamics, and the hadronic tensor,
\begin{equation}
W^{\mu\nu} = \frac{1}{4\pi} \sum_X (2\pi)^4 \delta^{(4)}(P+q-p_X)\,
\langle P|J^\mu(0)|X\rangle\,
\langle X|J^\nu(0)|P\rangle
\label{eq:hadtensor}
\end{equation}
contains all information about the strong-interaction dynamics of the target, with the sum extending over all possible hadronic final states~$X$.
The hadronic tensor is constrained by Lorentz covariance, electromagnetic current conservation, time-reversal invariance, and (for electromagnetic interactions) parity conservation. For an unpolarized target it can be parametrized in terms of only two independent scalar structure functions,
\begin{align}
W^{\mu\nu} 
&= \bigg( -g^{\mu\nu} + \frac{q^\mu q^\nu}{q^2} \bigg)\, F_1(x_B,Q^2)
+ \frac{1}{P\cdot q} 
  \bigg( P^\mu - \frac{P\cdot q}{q^2}q^\mu \bigg)\,
  \bigg( P^\nu - \frac{P\cdot q}{q^2}q^\nu \bigg)\, F_2(x_B,Q^2).
\label{eq:Wmunu}
\end{align}
The structure functions $F_1$ and $F_2$ completely characterize the response of an unpolarized nucleon to an electromagnetic probe, and are traditionally written as functions of the Bjorken variable and the momentum transfer $Q^2$.
It is often convenient to also define the longitudinal structure function,
\begin{equation}
F_L(x_B,Q^2) = \rho^2\,F_2(x_B,Q^2) - 2x_B F_1(x_B,Q^2),
\label{eq:FL}
\end{equation}
where 
\begin{equation}
\rho^2 = 1+\frac{4 M^2 x_B^2}{Q^2},
\label{eq:rho2}
\end{equation}
which measures the response of the nucleon to longitudinally polarized virtual photons. While $F_2$ receives contributions from both transverse and longitudinal photon polarizations, $F_L$ isolates the longitudinal component and is therefore particularly sensitive to QCD dynamics beyond the parton model. In the naive parton model, in which the virtual photon scatters incoherently from free spin-$\frac12$ quarks, one obtains the Callan--Gross relation, $F_2(x_B) = 2 x_B F_1(x_B)$, or $F_L(x_B) = 0$. The vanishing of the longitudinal structure function reflects the spin-$\frac12$ nature of the quark constituents, and was one of the earliest successes of the parton model. In QCD, however, perturbative gluon radiation, finite target-mass effects, and higher-twist contributions can generate a nonzero $F_L$. Measurements of $F_L$ therefore provide constraints on the gluon distribution as well as on power corrections at moderate values of $Q^2$.

For polarized scattering, the antisymmetric part of the hadronic tensor, $W_A^{\mu\nu}$, introduces two additional structure functions, $g_1(x_B,Q^2)$ and $g_2(x_B,Q^2)$,
\begin{align}
W_A^{\mu\nu}
&=
i\epsilon^{\mu\nu\alpha\beta}
\frac{q_\alpha}{P\cdot q}
\left[
S_\beta\, g_1(x_B,Q^2)
+
\left(
S_\beta
-
\frac{S\cdot q}{P\cdot q}\,P_\beta
\right)
g_2(x_B,Q^2)
\right].
\end{align}
Here $S^\mu$ is the nucleon spin four-vector. The structure function $g_1$ is dominated at large $Q^2$ by its twist-two contribution, which is related to helicity-dependent PDFs, but it also receives higher-twist corrections suppressed by powers of $1/Q^2$. In contrast, $g_2$ contains both a twist-two contribution, determined by $g_1$ through the Wandzura--Wilczek relation, and a genuine twist-three contribution associated with quark--gluon correlations. These twist-three terms are not suppressed by an additional power of $1/Q^2$ relative to the twist-two contribution to $g_2$, although their contribution to particular observables may be accompanied by kinematic factors involving the hard scale.

The asymptotic regime of DIS is defined by the Bjorken limit, $Q^2,\ \nu\rightarrow\infty$ and $x_B$ fixed, in which the structure functions become approximately independent of $Q^2$. This phenomenon, known as Bjorken scaling, provided one of the first indications that the nucleon contains pointlike constituents. Within QCD, approximate scaling is understood as a consequence of asymptotic freedom, while the observed logarithmic scaling violations arise from perturbative gluon radiation and are quantitatively described by the DGLAP evolution equations. The leading-twist description therefore becomes increasingly accurate as the momentum transfer grows, with corrections suppressed by inverse powers of $Q^2$. At finite $Q^2$, however, several distinct classes of power corrections become important. Kinematic effects associated with the finite nucleon mass generate target-mass corrections proportional to $x_B^2 M^2/Q^2$~[\cite{Nachtmann:1973mr, Georgi:1976ve}], while dynamical higher-twist contributions arise from coherent interactions involving two or more partons. Additional corrections originate from heavy-quark masses, threshold effects, and nuclear dynamics in scattering from nuclei. Separating these various contributions has become an important aspect of modern phenomenological analyses, particularly in the large-$x_B$ region where they may become numerically comparable to the leading-twist contribution.

It is also useful to distinguish between different regions of the hadronic final-state invariant mass. For $W \gtrsim 2~{\rm GeV}$, the scattering process is dominated by the deep-inelastic continuum, where the leading-twist description is expected to provide an accurate representation of the data. At lower invariant masses, the reaction proceeds through the excitation of nucleon resonances, giving rise to a rich spectrum of hadronic states. Although resonance production is intrinsically nonperturbative, it exhibits the remarkable phenomenon of quark--hadron duality, whereby suitably averaged resonance-region structure functions closely follow the scaling behavior observed in the deep-inelastic continuum. This observation suggests that the transition between hadronic and partonic degrees of freedom is considerably smoother than might naively be expected and provides important insight into the interplay between perturbative and nonperturbative QCD. The theoretical framework for describing the scaling regime is provided by the QCD factorization theorem, which enables the systematic separation of short- and long-distance dynamics. At sufficiently large $Q^2$, the structure functions factorize into universal PDFs and perturbatively calculable hard-scattering coefficient functions, providing the foundation for modern analyses of DIS. This leading-twist description establishes the baseline against which power-suppressed effects are identified. We therefore begin by reviewing the QCD factorization formalism at leading twist before considering the systematic organization of the power corrections.

\subsection{Leading-Twist Factorization}
\label{sec:factorization}

The approximate scaling of the structure functions observed in the Bjorken limit is naturally explained within perturbative QCD through the QCD factorization theorem~[\cite{Collins:1989gx, Collins:2011zzd}]. At sufficiently large momentum transfers, $Q^2 \gg \Lambda_{\rm QCD}^2$, the hard interaction between the virtual photon and an individual parton occurs over distances much smaller than the characteristic hadronic scale, allowing the short-distance dynamics to be separated systematically from the long-distance structure of the nucleon. For inclusive DIS, the appropriate framework is collinear factorization, in which the active partons are treated as carrying only longitudinal momentum, with their transverse momenta integrated over and absorbed into universal PDFs.

The resulting factorization formula for the leading-twist contribution to the structure function $F_i$ is valid up to corrections suppressed by powers of the hard scale, and may be written schematically as [\cite{Blumlein:2012bf}]
\begin{equation}
F_i^{\rm LT}(x_B,Q^2) = \sum_{a=q,\bar q,g}
\left[ C_{i,a}\otimes f_a \right](x_B,Q^2),
\label{eq:factorization}
\end{equation}
where $C_{i,a}$ is the perturbatively calculable Wilson coefficient, $f_a(x,\mu_F^2)$ is the corresponding PDF evaluated at the factorization scale $\mu_F$, and $\otimes$ denotes the convolution
\begin{equation}
\left[ C\otimes f \right](x_B)
= \int_{x_B}^{1} \frac{dx}{x}\,
C\!\left( \frac{x_B}{x},\frac{Q^2}{\mu_F^2},\alpha_s(\mu_R^2) \right)
f(x,\mu_F^2),
\label{eq:convolution}
\end{equation}
where $\mu_R$ is the renormalization scale at which the strong coupling $\alpha_s$ is defined. Although physical observables are independent of the choices of $\mu_F$ and $\mu_R$ to all orders in perturbation theory, residual scale dependence remains at finite order and provides an estimate of the theoretical uncertainty associated with the perturbative calculation.

At leading order, the virtual photon scatters elastically from an
individual quark, so that the coefficient function is proportional to
$\delta(1-x_B/x)$. The structure function $F_2$ then reduces to the
familiar quark-parton model expression,
\begin{equation}
F_2^{\rm LO}(x_B,Q^2) = x_B
\sum_q e_q^2 \left[ q(x_B,Q^2) + \bar q(x_B,Q^2) \right].
\label{eq:F2LO}
\end{equation}
Beyond leading order, gluon radiation and $q\bar q$ pair production generate nontrivial coefficient functions, while the gluon distribution enters through higher-order subprocesses.

The scale dependence of the PDFs is governed by the DGLAP evolution equations~[\cite{GribovLipatov1972, AltarelliParisi1977, Dokshitzer1977}],
\begin{equation}
\frac{\partial f_a(x,\mu_F^2)}{\partial\ln\mu_F^2}
= \sum_b P_{ab} \otimes f_b, 
\label{eq:DGLAP}
\end{equation}
where $P_{ab}$ are the splitting functions describing the probability for one parton to radiate another. These equations resum the logarithmic dependence on the hard scale arising from collinear parton radiation and provide the quantitative description of scaling violations observed in DIS. Phenomenologically, the evolution equations establish the predictive power of perturbative QCD by relating measurements performed at different values of $Q^2$. Both the coefficient functions and the splitting functions are known to high orders in perturbation theory, with calculations currently available through next-to-next-to-next-to-leading order (N$^3$LO) for many inclusive observables. As a consequence, the theoretical uncertainties associated with the leading-twist description have been substantially reduced, placing increasing emphasis on understanding the power-suppressed corrections that become important at moderate values of $Q^2$.

The factorized expression in Eq.~(\ref{eq:factorization}) represents only the leading term in an expansion in inverse powers of the hard scale. More generally, one can write
\begin{equation}
F_i(x_B,Q^2)
= F_i^{\rm LT}(x_B,Q^2)
+ {\cal O}\!\left(\frac{M^2}{Q^2},\frac{\Lambda_{\rm QCD}^2}{Q^2}\right),
\label{eq:LTpower}
\end{equation}
where the first correction arises from the finite target mass, and the second represents dynamical contributions associated with multiparton correlations, which are controlled by the QCD scale parameter $\Lambda_{\rm QCD}$. At large values of $x_B$, threshold logarithms, higher-order perturbative corrections, and power corrections all become numerically significant, making a careful treatment of the leading-twist contribution essential before genuine higher-twist effects can be reliably extracted from experimental data.

\subsection{Operator Product Expansion and Moments}
\label{sec:OPE}

The leading-twist factorization theorem provides the asymptotic description of DIS in the Bjorken limit, where power corrections suppressed by inverse powers of $Q^2$ may be neglected. A systematic framework for organizing these corrections is provided by the OPE~[\cite{Wilson:1969zs}], which exploits the fact that the forward virtual Compton amplitude is dominated by light-cone distances, $z^2\rightarrow0$, at large momentum transfer. The time-ordered product of two electromagnetic currents, $J_\mu$, may therefore be expanded as a series of local operators,
\begin{equation}
i\int d^4z\,e^{iq\cdot z}
T\!\left[
J_\mu(z)J_\nu(0)
\right]
=
\sum_{n,\tau}
C_{\mu\nu}^{(n,\tau)}(q,\mu)\,
O^{(n,\tau)}(\mu),
\label{eq:OPE}
\end{equation}
where the Wilson coefficient functions $C_{\mu\nu}^{(n,\tau)}$ are perturbatively calculable and describe the short-distance dynamics, while the local operators $O^{(n,\tau)}$ encode the long-distance structure of the nucleon. The operators are classified according to their Lorentz spin $n$ and twist,
\begin{equation}
\tau=d-n,
\label{eq:twist}
\end{equation}
where $d$ is the canonical mass dimension of the operator. Operators of the lowest twist dominate in the Bjorken limit, while operators of progressively higher twist generate contributions suppressed by successive powers of $1/Q^2$.

\begin{figure}[t]
\centering
\includegraphics[width=0.82\textwidth]{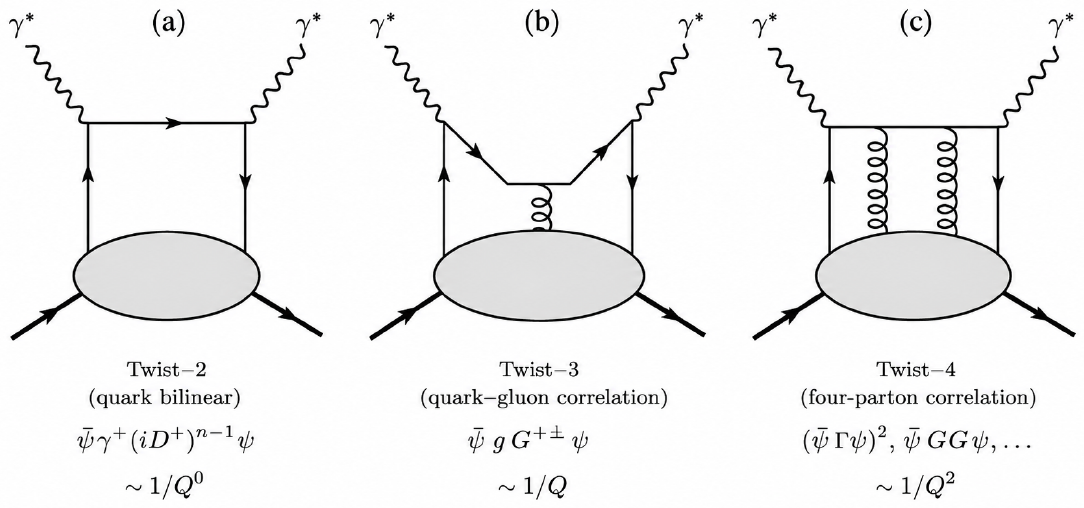}
\vspace*{-13cm}
\caption{Representative operator topologies contributing to inclusive DIS: (a) twist-2 quark bilinear operator corresponding to scattering from a single parton, (b) twist-3 quark--gluon correlation operator, and (c) twist-4 operator involving four-parton correlations.}
\label{fig:twist}
\end{figure}

For the spin-independent structure functions, the leading contributions arise from the symmetric, traceless twist-two quark and gluon operators,
\begin{align}
O_q^{\mu_1\cdots\mu_n}
&=
\bar\psi
\gamma^{(\mu_1}
iD^{\mu_2}
\cdots
iD^{\mu_n)}
\psi
-
{\rm traces},
\\
O_g^{\mu_1\cdots\mu_n}
&=
F^{(\mu_1\alpha}
iD^{\mu_2}
\cdots
iD^{\mu_{n-1}}
F^{\mu_n)}_{\phantom{\mu_n}\alpha}
-
{\rm traces},
\end{align}
whose forward matrix elements are related to Mellin moments of the leading-twist quark and gluon PDFs. Higher-twist operators contain additional parton fields or different Lorentz projections and therefore probe coherent multiparton correlations inside the nucleon. Representative examples include the twist-three quark--gluon operator
\begin{equation}
{\cal O}_{qG}^{\sigma\{\mu_1\cdots\mu_n\}}
=
\bar\psi\,
g\widetilde G^{\sigma(\mu_1}
\gamma^{\mu_2}
iD^{\mu_3}
\cdots
iD^{\mu_n)}
\psi
-
{\rm traces},
\end{equation}
which contributes to the genuine twist-three component of the polarized structure function $g_2$, and the twist-four quark--gluon and four-quark operators
\begin{align}
O_{qG}^{\mu\nu}
&=
g\,
\bar q\,
G^{\alpha\{\mu}
\gamma_\alpha
iD^{\nu\}}
q
-
{\rm traces},
\\
O_{4q}^{\mu\nu}
&=
g^2
\left(
\bar q\gamma^{\{\mu}t^Aq
\right)
\left(
\bar q\gamma^{\nu\}}t^Aq
\right)
-
{\rm traces},
\end{align}
which generate the leading dynamical $1/Q^2$ corrections to unpolarized DIS~[\cite{Jaffe:1991kp, Jaffe:1991ra, Shuryak:1981kj, Shuryak:1982dp}]. Although the choice of operator basis is not unique because operators with identical quantum numbers mix under renormalization and may be related through the equations of motion and total derivatives, physical structure functions are independent of the basis adopted. Representative leading- and higher-twist operator topologies are shown in Fig.~\ref{fig:twist}. The twist-two operator corresponds to incoherent scattering from a single parton, whereas the twist-three and twist-four operators involve explicit quark--gluon and multiparton correlations.

The connection between these local operators and experimentally measurable observables is established through moments of the structure functions. As a representative example, consider the Cornwall--Norton moments of the unpolarized structure function $F_2$,
\begin{equation}
M_2^{(n)}(Q^2)
= \int_0^1 dx_B\,x_B^{\,n-2}F_2(x_B,Q^2),
\label{eq:CNmoment}
\end{equation}
with analogous definitions for the remaining unpolarized and polarized structure functions. At leading twist, the moments are directly related to Mellin moments of the PDFs, while the Wilson coefficients become ordinary products in moment space.

The OPE organizes the moments as an expansion in operators of increasing twist,
\begin{equation}
M_2^{(n)}(Q^2)
=
C_2^{(2,n)}(Q^2,\mu^2)\,
A^{(2,n)}(\mu)
+
\frac{1}{Q^2}
\sum_i
C_{2,i}^{(4,n)}(Q^2,\mu^2)\,
A_i^{(4,n)}(\mu)
+
{\cal O}\!\left(\frac{1}{Q^4}\right),
\label{eq:HTgeneral}
\end{equation}
where the Wilson coefficients are perturbatively calculable, while $A_i^{(\tau,n)}$ are reduced nucleon matrix elements of local operators of twist $\tau$. Forward matrix elements of symmetric, traceless operators are parameterized by
\begin{equation}
\langle P|
O_{\{\mu_1\cdots\mu_n\}}^{(\tau)}(\mu)
|P\rangle
=
2A_{n,O}^{(\tau)}(\mu)
\left(
P_{\mu_1}\cdots P_{\mu_n}
-
{\rm traces}
\right),
\label{eq:HTME}
\end{equation}
where the reduced matrix elements contain the nonperturbative information associated with the corresponding local operators.

For many phenomenological applications it is convenient to write the moment in terms of an effective twist expansion,
\begin{equation}
M_2^{(n)}(Q^2)
=
M_{2,\rm LT}^{(n)}(Q^2)
+
\frac{1}{Q^2}
A_2^{(4,n)}(Q^2)
+\cdots ,
\label{eq:effectiveHT}
\end{equation}
where
\begin{equation}
A_2^{(4,n)}(Q^2)
=
\sum_i
C_{2,i}^{(4,n)}(Q^2,\mu^2)\,
A_i^{(4,n)}(\mu),
\label{eq:A4def}
\end{equation}
represents the appropriate observable-dependent combinations of Wilson coefficients and twist-four reduced matrix elements. The weighting factor $x_B^{\,n-2}$ in Eq.~(\ref{eq:CNmoment}) increasingly suppresses the small-$x_B$ region with increasing moment order, making higher moments particularly sensitive to the large-$x_B$ region where genuine higher-twist effects are expected to be most important. Moments therefore provide a natural framework for comparing experiment with lattice QCD calculations of local operator matrix elements and for studying quark--hadron duality.

A systematic classification of local QCD operators was established by~[\cite{Jaffe:1981td}] according to their twist, spin, and Lorentz transformation properties, providing a rigorous operator basis for the OPE description of DIS. Using equations of motion and symmetry constraints, they identified the independent quark, gluon, and mixed quark--gluon operators contributing at each twist and clarified the role of operator mixing under renormalization. Their formalism further clarified the connection between the twist expansion of DIS moments and the underlying gauge-invariant local operators of QCD, providing the foundation for subsequent analyses of higher-twist anomalous dimensions, renormalization-group evolution, and phenomenological studies of power corrections. Together with the contemporaneous work of~[\cite{Shuryak:1981kj, Shuryak:1982dp}], the Jaffe--Soldate analysis established the modern operator-based framework for higher-twist effects in DIS.

In addition to the full Cornwall--Norton moments, it is often useful to consider truncated moments,
\begin{equation}
M_2^{(n)}
(x_{\rm min},x_{\rm max};Q^2)
=
\int_{x_{\rm min}}^{x_{\rm max}}
dx_B\,
x_B^{\,n-2}
F_2(x_B,Q^2),
\label{eq:truncmoment}
\end{equation}
which isolate restricted regions of phase space, such as the resonance region. Although truncated moments no longer possess the direct local-operator interpretation of the full Mellin moments, they satisfy closed evolution equations~[\cite{Forte:1999tx, Kotlorz:2004zy}] and are useful for quantitative studies of quark--hadron duality (see below).

\subsection{Diagrammatic Approach}
\label{sec:EFP}

Although the OPE provides the theoretical foundation for the analysis of higher-twist effects in inclusive DIS, its applicability is limited to observables that can be expressed in terms of matrix elements of local operators between single-hadron states. For hard-scattering processes that involve two or more identified hadrons, such as SIDIS, Drell--Yan lepton-pair production, or inclusive hadron production in hadronic collisions, the OPE no longer provides a general proof of factorization, even though the underlying short-distance current product may still be expanded in terms of local operators. For such processes the appropriate theoretical framework is provided by a diagrammatic analysis of QCD amplitudes in momentum space, which allows the systematic factorization of perturbative hard-scattering coefficients from universal nonperturbative multiparton correlation functions.

In general, any hard-scattering observable characterized by a large momentum transfer $Q$ can be expanded in inverse powers of $Q$, with the accompanying dimensionful scales supplied by nonperturbative hadronic matrix elements. For observables involving a single identified hadron, such as inclusive DIS, the factorization into perturbatively calculable short-distance functions and universal long-distance correlation functions is established by the OPE, which expresses the moments of the structure functions in terms of matrix elements of local operators of specific twist. An equivalent description can also be formulated diagrammatically through the momentum-space approach of Ellis, Furmanski, and Petronzio (EFP), which reproduces the same power expansion in terms of hard-scattering coefficient functions convoluted with gauge-invariant multiparton correlation functions~[\cite{Ellis:1982wd, Ellis:1982cd}]. In this framework, Feynman diagrams can be systematically classified according to their scaling with the hard scale $Q$, and the DIS hadronic tensor can be expanded as
\begin{equation}
W_{\mu\nu} 
= H_{\mu\nu}^{(2)} \otimes f
+ \frac{1}{Q^2}
\left(
H_{\mu\nu,q}^{(4)} \otimes T_q + H_{\mu\nu,g}^{(4)} \otimes T_g
\right)
+ \cdots,
\label{eq:EFP}
\end{equation}
where $H_{\mu\nu}^{(2)}$ and $H_{\mu\nu,i}^{(4)}\ (i=q,g)$ are perturbatively calculable hard-scattering coefficient functions, $f$ is the leading-twist PDF, and $T_q$ and $T_g$ are universal twist-four multiparton correlation functions, respectively. In contrast to the leading-twist handbag approximation, the power-suppressed terms arise from coherent interactions involving additional soft partons, thereby providing a direct physical interpretation of higher twists as manifestations of multiparton dynamics within the nucleon.

Building on the EFP formalism, Qiu and Sterman generalized the diagrammatic approach to hadronic scattering and demonstrated that both the leading-power and first power-suppressed contributions to processes with two identified hadrons can be factorized in terms of universal, gauge-invariant multiparton correlation functions and perturbatively calculable hard-scattering coefficients~[\cite{ Qiu:1990xxa, Qiu:1990xy, Qiu:1991pp, Qiu:1991wg, Qiu:1998ia}]. This work established that coherent multiple-parton interactions can be treated systematically within collinear factorization, thereby extending the diagrammatic factorization program to hard-scattering processes involving more than one identified hadron.

A central ingredient of the Qiu--Sterman formalism is the appearance of
nonlocal quark--gluon correlation functions, which may be written
schematically as
\begin{equation}
T_q(x_1,x_2)
\propto
\int dy_1^-\,dy_2^-\, e^{ix_1P^+y_1^-} e^{i(x_2-x_1)P^+y_2^-}\,
\langle P|
\bar{\psi}(0)\, \gamma^+\, F^{+\alpha}(y_2^-)\, \psi(y_1^-)
|P\rangle ,
\label{eq:QScorr}
\end{equation}
where gauge links connecting the fields, necessary to ensure gauge invariance, are implied. Here, $x_1$ and $x_2$ denote the light-cone momentum fractions carried by the active quark before and after its interaction with the gluon field, respectively, with $x_2-x_1$ the momentum fraction carried by the coherent gluon. The coordinates $y_1^-$ and $y_2^-$ specify the corresponding light-cone separations of the quark and gluon fields, respectively. Unlike the leading-twist PDFs, which depend on a single parton momentum fraction, the quark--gluon correlation function depends on two independent momentum fractions, reflecting the coherent participation of both a quark and a gluon in the hard scattering. These correlation functions constitute the fundamental nonperturbative quantities entering higher-twist factorization and provide a physical description of coherent multiparton interactions inside hadrons.

At present, however, no general proof of collinear factorization has been established to all orders in the $1/Q$ expansion for observables involving two or more identified hadrons. Beyond the first subleading power, the number of independent multiparton correlation functions increases rapidly, while long-distance soft interactions and more intricate color correlations considerably complicate the factorization analysis. Nevertheless, the work of Qiu and Sterman established the theoretical foundation upon which much of the modern higher-twist formalism has been built, and was subsequently extended to describe coherent multiple scattering in nuclei, nuclear-enhanced power corrections, transverse-momentum broadening, as well as the collinear twist-three formalism for spin-dependent observables. Within this framework, the Efremov--Teryaev--Qiu--Sterman (ETQS) quark--gluon correlation function~[\cite{Efremov:1981sh, Efremov:1984ip, Qiu:1991pp, Qiu:1991wg}] plays a central role in generating large transverse single-spin asymmetries through coherent quark--gluon interactions, while its scale dependence is determined by perturbatively calculable evolution equations~[\cite{Kang:2008ey}].

Complementing the factorization program of EFP and Qiu--Sterman, [\cite{Balitsky:1987bk}] developed a gauge-invariant light-ray operator formalism that provides the general framework for the renormalization-group evolution and operator mixing of higher-twist operators. Their work established the higher-twist analog of DGLAP evolution, while later calculations by Kang and Qiu derived the explicit evolution equations for the ETQS quark--gluon correlation functions used in modern twist-three phenomenology~[\cite{Kang:2008ey}]. The diagrammatic factorization approach pioneered by EFP and extended by Qiu and Sterman, and the operator-based evolution formalism developed by Balitsky and Braun, provide the complementary ingredients required for a complete description of higher-twist QCD dynamics: the former establishes the factorization of hard-scattering cross sections in terms of universal multiparton correlation functions, while the latter determines their scale dependence. Combined with the OPE treatment of inclusive DIS, these developments have established a comprehensive theoretical framework for systematically incorporating higher-twist effects into precision QCD phenomenology.

\subsection{Target Mass Corrections}
\label{sec:TMC}

The OPE discussion in Sec.~\ref{sec:OPE} is strictly valid in the Bjorken limit, where the nucleon mass is neglected relative to the hard scale. At finite values of $Q^2$, however, the finite target mass gives rise to corrections proportional to powers of $x_B^2 M^2/Q^2$. These target mass corrections (TMCs) become increasingly important at large $x_B$, where they can significantly modify the measured structure functions even when $Q^2$ is of the order of several GeV$^2$.

Unlike genuine higher-twist effects, TMCs do not represent new information about the internal structure of the nucleon. Rather, they arise entirely from the exact kinematics of scattering from a target of finite mass and therefore constitute kinematic corrections to the leading-twist contribution. Their magnitude is often comparable to that of genuine higher-twist effects in the kinematic region explored by fixed-target experiments, making their consistent treatment essential in any extraction of higher-twist contributions from data.

The finite target mass modifies the relation between the Bjorken scaling variable $x_B$ and the light-cone momentum fraction carried by the struck parton. The appropriate scaling variable is the Nachtmann variable~[\cite{Bhaumik:1971, Nachtmann:1973mr}],
\begin{equation}
\xi = \frac{2x_B}{1+\rho},
\label{eq:xi}
\end{equation}
which reduces to the Bjorken variable in the asymptotic limit, $\xi\rightarrow x_B$ as $Q^2\rightarrow\infty$.

Within the OPE, TMCs arise entirely from trace terms in the matrix elements of twist-two operators and were first derived by Georgi and Politzer~[\cite{Georgi:1976ve}]. For the $F_2$ structure function the result may be written as
\begin{align}
F_2^{\rm OPE}(x_B,Q^2)
&= \frac{(1+\rho)^2}{4\rho^3}\, F_2^{(0)}(\xi,Q^2)
+ \frac{3x_B(\rho^2-1)}{2\rho^4}
\left[ h_2(\xi,Q^2) + \frac{\rho^2-1}{2x_B\rho} g_2(\xi,Q^2) \right],
\label{eq:F2OPE}
\end{align}
where $F_2^{(0)}$ denotes the massless structure function and
\begin{align}
h_2(\xi,Q^2)
&= \int_\xi^1 du\, \frac{F_2^{(0)}(u,Q^2)}{u^2},
\\
g_2(\xi,Q^2)
&= \int_\xi^1 dv\, \int_v^1 du\, \frac{F_2^{(0)}(u,Q^2)}{u^2}.
\end{align}

An alternative formulation is provided by collinear factorization, originally developed by~[\cite{Aivazis:1993kh, Aivazis:1993pi}] and subsequently clarified for inclusive DIS with exact external kinematics by~[\cite{Moffat:2019jto}]. In this approach the factorization derivation is performed without making a massless-target approximation, so that the Nachtmann variable $\xi$ appears naturally as the scaling variable. For the $F_2$ structure function, one has
\begin{equation}
F_2^{\rm CF}(x_B,Q^2)
=
\frac{x_B}{\xi\,\rho^2}\,
F_2^{(0)}(\xi,Q^2)
+
{\cal O}\!\left(\frac{m^2}{Q^2}\right),
\label{eq:F2TMC_CF}
\end{equation}
where $m^2$ denotes intrinsic partonic scales, such as parton virtuality or transverse momentum, that are neglected in the leading-power collinear approximation.

The Cornwall--Norton moments introduced in the previous subsection receive kinematic contributions from trace terms that mix operators of different spin. Nachtmann showed that this mixing can be removed by constructing moments that project operators of definite spin. For the $F_2$ structure function the Nachtmann moments are
\begin{equation}
M_{2}^{(n)\,N}(Q^2)
=
\int_0^1
dx_B\,
\frac{\xi^{\,n+1}}{x_B^3}
\,
\frac{
3+3(n+1)\rho+n(n+2)\rho^2
}
{(n+2)(n+3)}
F_2(x_B,Q^2),
\label{eq:Nachtmann}
\end{equation}
which reduce to the Cornwall--Norton moments in the Bjorken limit. By construction, the Nachtmann moments receive contributions only from operators of spin $n$, allowing TMCs to be separated from genuine higher-twist effects within the OPE. Although both TMCs and higher-twist effects contribute corrections that scale as inverse powers of $Q^2$, their physical origins are entirely different. TMCs arise solely from the finite target mass and are contained within the twist-two contribution, whereas genuine higher-twist effects originate from matrix elements of operators with twist greater than two. In phenomenological analyses it is therefore essential to incorporate TMCs into the leading-twist baseline befored attributing any remaining deviations from scaling to dynamical multiparton correlations.

The formalism developed in this section provides the theoretical framework for understanding the scaling behavior of DIS structure functions and the origin of power corrections at finite $Q^2$. While the OPE predicts that higher-twist contributions are suppressed by increasing powers of $1/Q^2$, an important question is how these effects manifest themselves in experimentally measured structure functions. One of the most remarkable observations is that, even in the nucleon resonance region where the scattering process is dominated by individual hadronic excitations, suitably averaged structure functions closely follow the scaling behavior expected from the leading-twist description. This phenomenon, known as quark--hadron duality, provides a unique window on the transition between quark and hadron degrees of freedom and offers important insights into the interplay between perturbative and nonperturbative QCD. We review the experimental evidence and theoretical understanding of quark--hadron duality in the following section.

\section{Quark--Hadron Duality}
\label{sec:duality}

Quark--hadron duality is one of the most striking manifestations of the interplay between perturbative and nonperturbative dynamics in QCD. It expresses the empirical observation that structure functions measured in the nucleon resonance region, when suitably averaged, closely follow the scaling behavior observed in DIS, despite the very different physical descriptions of the two regimes. Since its discovery more than five decades ago, quark--hadron duality has evolved from an intriguing experimental observation into a tool for studying the transition between hadronic and partonic degrees of freedom. It has also provided important insight into the role of higher-twist effects, the onset of scaling, and the applicability of perturbative QCD at low $Q^2$.
This section reviews both the experimental evidence for duality and its modern theoretical interpretation within QCD. We begin in Sec.~\ref{sec:duality_history} with a brief historical overview, describing the original observations of Bloom and Gilman and the development of the concept prior to QCD. Section~\ref{sec:duality_qcd} then discusses the interpretation of duality in terms of the OPE and the twist expansion, emphasizing the role of higher-twist contributions and their apparent suppression upon averaging over resonance regions, while experimental tests of duality are summarized in Sec.~\ref{sec:duality_exp}.

\subsection{Historical Development}
\label{sec:duality_history}

One of the most remarkable phenomena observed in inclusive lepton--nucleon scattering is the intimate connection between the nucleon resonance and DIS regions. Rather than representing distinct dynamical regimes, measurements have shown that suitably averaged resonance structure functions closely follow the scaling curves measured at much higher energies. This phenomenon provides an important link between the perturbative quark-gluon description of QCD and the nonperturbative physics of hadronic resonances.

The first clear evidence for this behavior was reported by Bloom and Gilman [\cite{Bloom:1970xb, Bloom:1971ye}], who observed that the prominent resonance peaks in the DIS cross section at low values of the hadronic invariant mass $W$ did not fluctuate randomly about the scaling curve measured in the DIS region, but rather appeared to oscillate around a universal curve that was nearly independent of $Q^2$. Even more surprisingly, averaging the resonance contributions over an interval in $W$ reproduced the scaling function measured at substantially larger values of $Q^2$, where individual resonances are no longer resolved.

The observation of resonance scaling was unexpected because the resonance and DIS regions are traditionally described using very different degrees of freedom. Resonance production is naturally understood in terms of hadronic excitations of the nucleon, whereas the parton model interprets DIS as incoherent scattering from nearly free quarks. The close numerical agreement between the averaged resonance cross section and the scaling structure function therefore suggested that the underlying quark dynamics somehow survive the complicated process of hadronization.

\begin{figure}[t]
\centering
\includegraphics[width=0.75\textwidth]{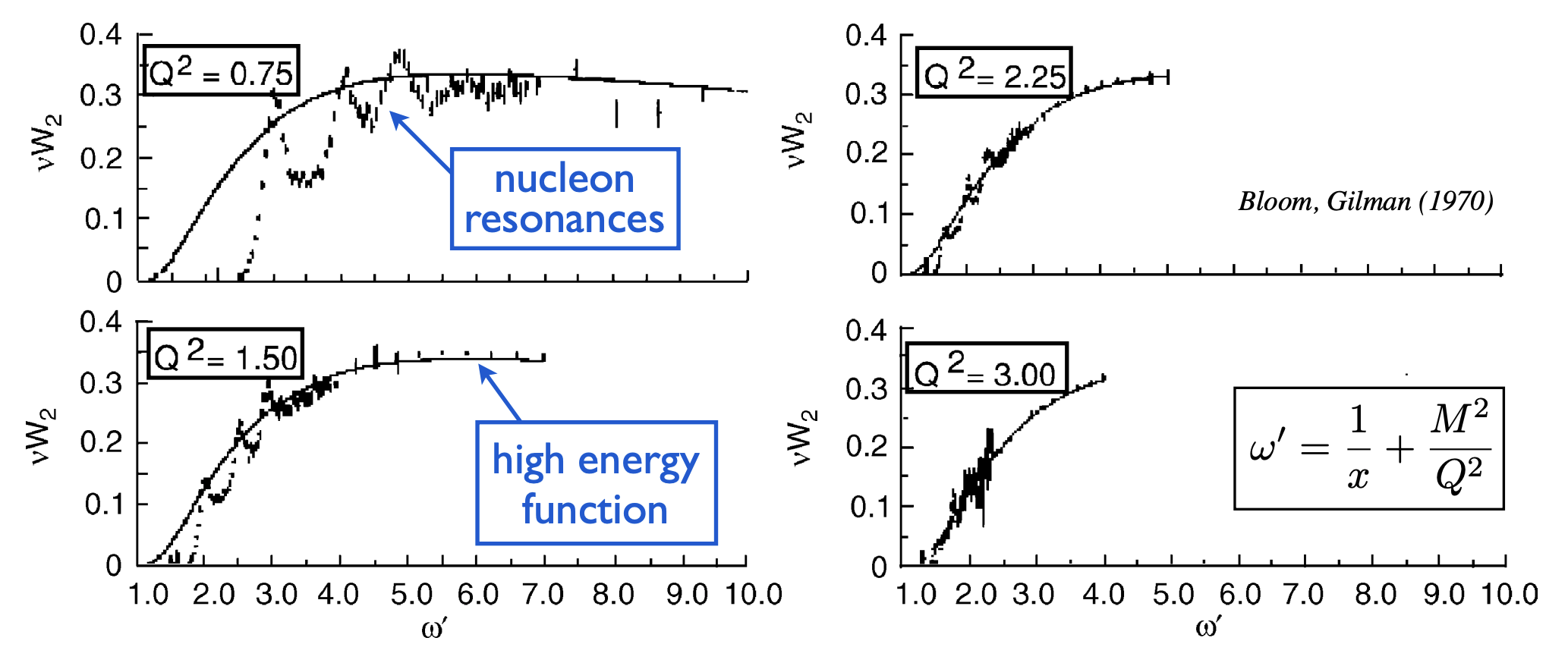}
\caption{Representative measurements of the proton structure function $F_2$ versus the Bloom-Gilman scaling variable $\omega' = 1 + W^2/Q^2 = 1/x_B + M^2/Q^2$ in the resonance region compared with the scaling curve measured at higher values of $Q^2$, illustrating the original Bloom--Gilman observation of quark--hadron duality. Adapted from~[\cite{Bloom:1970xb, Bloom:1971ye}].}
\label{fig:dualityBG}
\end{figure}

The idea that averages over hadronic states can reproduce quark-level predictions predates QCD. Similar concepts had been developed in the context of finite-energy sum rules and dual resonance models, where cross sections could be described either by a sum over $s$-channel resonances or by $t$-channel Regge exchanges, with both descriptions yielding equivalent results after suitable averaging~[\cite{Dolen:1967jr}]. The Bloom--Gilman observations, illustrated in Fig.~\ref{fig:dualityBG}, provided one of the first experimental manifestations of the notion of duality in strong interactions.

The original experimental observations are commonly expressed in terms of integrals over restricted regions of the structure functions. For a given interval in $x_B$, or more commonly in $W$, the averaged resonance contribution is found to be approximately equal to the corresponding leading-twist prediction,
\begin{equation}
\int_{\Delta W}
dx_B\,
F_2(x_B,Q^2)
\approx
\int_{\Delta W}
dx_B\,
F_2^{\rm LT}(x_B,Q^2),
\label{eq:duality}
\end{equation}
where $F_2^{\rm LT}$ denotes the leading-twist structure function evolved to the same value of $Q^2$, and $\Delta W$ denotes the $x_B$ interval corresponding to a specific resonance or group of resonances. The observation that Eq.~(\ref{eq:duality}) is often satisfied to within $\approx 10$--$20\%$ for $Q^2\gtrsim1$--$2~{\rm GeV}^2$ was established only much later through high-precision measurements at Jefferson Lab, considerably extending the original SLAC studies.

Following the establishment of QCD as the theory of the strong interaction, quark--hadron duality acquired a natural theoretical interpretation through the OPE. From this perspective, the approximate equality between resonance averages and leading-twist structure functions reflects the fact that the net higher-twist contributions become surprisingly small after averaging over appropriate kinematic intervals, even though individual resonances are intrinsically nonperturbative. This modern interpretation has transformed duality from an empirical observation into a valuable probe of the transition between perturbative and nonperturbative QCD. In the following subsection we discuss the theoretical foundations of quark--hadron duality within QCD, emphasizing its interpretation in terms of the OPE and the twist expansion, and the role played by higher-twist contributions in understanding the onset of scaling.

\subsection{Duality in QCD}
\label{sec:duality_qcd}

The establishment of QCD as the fundamental theory of the strong interaction provided a natural theoretical framework for understanding quark--hadron duality. While the original Bloom--Gilman observations predated QCD, it was subsequently realized that duality can be interpreted in terms of the OPE, whereby the leading contribution arises from incoherent scattering from individual partons, while power-suppressed corrections encode multiparton correlations and other nonperturbative effects.

As discussed in Sec.~\ref{sec:OPE} above, the expansion in Eq.~(\ref{eq:effectiveHT}) represents a series of contributions to structure function moments of increasing twist. At asymptotically large values of $Q^2$, the power corrections vanish, and the moments are completely determined by the leading-twist PDFs. One might therefore expect the resonance region, where the cross section is dominated by narrow hadronic excitations, each of which is strongly dependent on $Q^2$, to differ substantially from the scaling structure functions measured in DIS. The empirical success of quark--hadron duality demonstrates that although individual resonances correspond to highly nonperturbative dynamics, their averaged contributions reproduce the leading-twist behavior with remarkable accuracy.

\begin{figure}[t]
\centering
\includegraphics[width=0.42\textwidth]{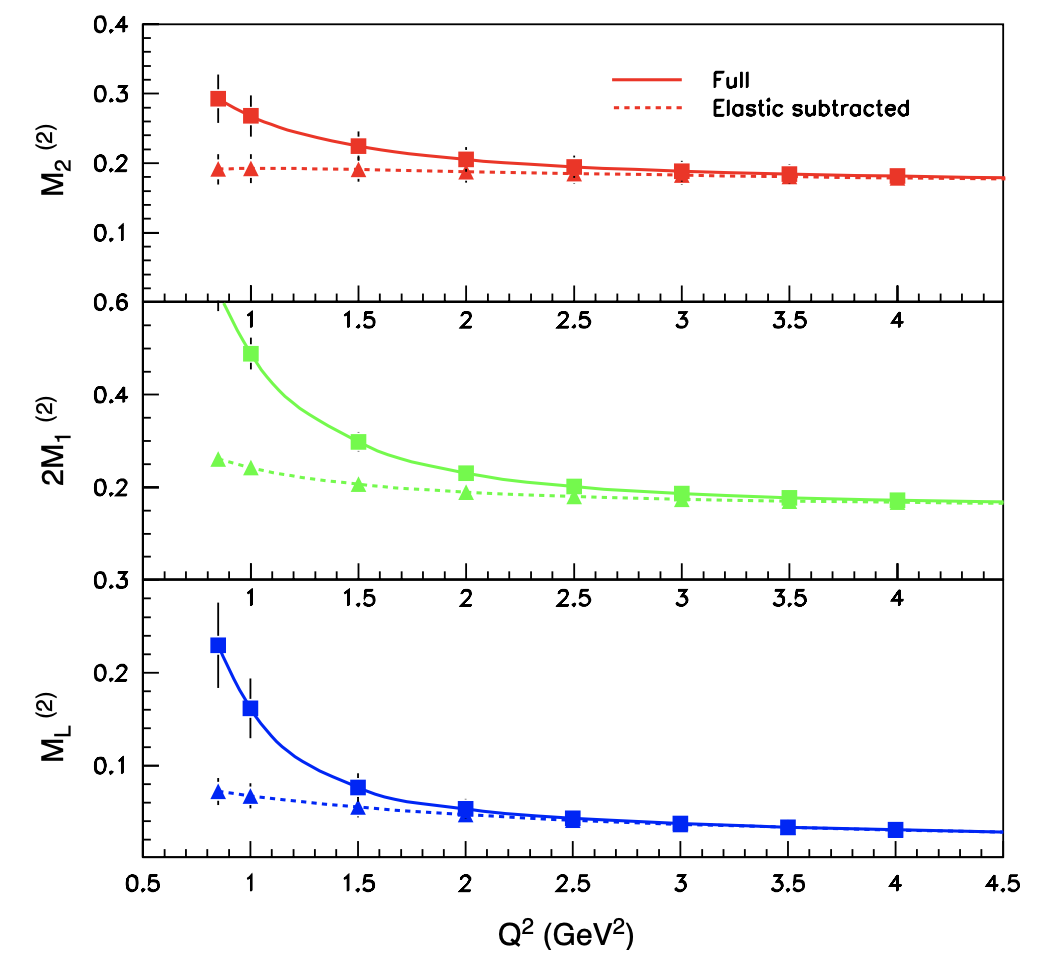}
\vspace*{-0.2cm}
\caption{Second ($n=2$) Cornwall–Norton moments of the proton $F_2$ (top), $2 x_B F_1$ (center) and $F_L$ (bottom) structure functions, evaluated from Jefferson Lab Hall~C data~[\cite{Liang:2003, JeffersonLabHallCE94-110:2004nsn}]. The total moments are connected by solid lines and elastic-subtracted moments by dashed lines to guide the eye. Adapted from~[\cite{Melnitchouk:2005zr}].}
\label{fig:momentsvsQ}
\end{figure}

The connection between duality and the OPE may be expressed schematically by considering the difference between the total structure function moment and its leading-twist approximation,
\begin{equation}
M_2^{(n)}(Q^2) - M_{2,\rm LT}^{(n)}(Q^2)
= \frac{1}{Q^2} A_2^{(4,n)}(Q^2)
+ {\cal O}\!\left(\frac{1}{Q^4}\right),
\label{eq:duality_difference}
\end{equation}
so that the averaged deviation from leading-twist behavior is suppressed by powers of $1/Q^2$. In this language, quark--hadron duality may be viewed as a direct manifestation of the smallness of the effective higher-twist contributions after averaging over resonance regions~[\cite{DeRujula:1976baf}]. Since the higher-twist contributions are suppressed by powers of $1/Q^2$, the resonance averages are expected to converge toward the leading-twist prediction with increasing $Q^2$. Remarkably, experimental measurements indicate that duality are already satisfied to good accuracy for $Q^2 \approx 1$--$2~{\rm GeV}^2$, where the expansion parameter $1/Q^2$ is not especially small. Figure~\ref{fig:momentsvsQ} illustrates the convergence of the moments of the proton structure functions with $Q^2$.

Although the OPE provides a compelling framework for interpreting quark--hadron duality, it does not by itself explain the dynamical origin of the cancellations among higher-twist contributions. This remains an active area of theoretical investigation involving lattice QCD, effective field theories, large-$N_c$ methods, and phenomenological models of resonance transition form factors. The theoretical picture outlined above is strongly supported by a broad range of experimental measurements. In the following subsection we review the evidence for quark--hadron duality obtained from measurements of structure functions over the resonance and DIS regions.

\subsection{Experimental Evidence}
\label{sec:duality_exp}

While the early SLAC measurements~[\cite{Bloom:1970xb, Bloom:1971ye}] established the existence of duality in the proton structure function $F_2$ (see Fig.~\ref{fig:dualityBG}), subsequent experiments extended these investigations over a much broader range of kinematics, to polarized scattering, neutron and nuclear targets, and neutrino interactions~[\cite{Melnitchouk:2005zr}]. The availability of high-luminosity electron beams and precision detectors, particularly at Jefferson Lab, has enabled systematic tests of duality to be performed with unprecedented accuracy~[\cite{Niculescu:2000tj, Niculescu:2000tk, Malace:2009kw}].

\begin{figure}[t]
\centering
\includegraphics[width=0.4\textwidth]{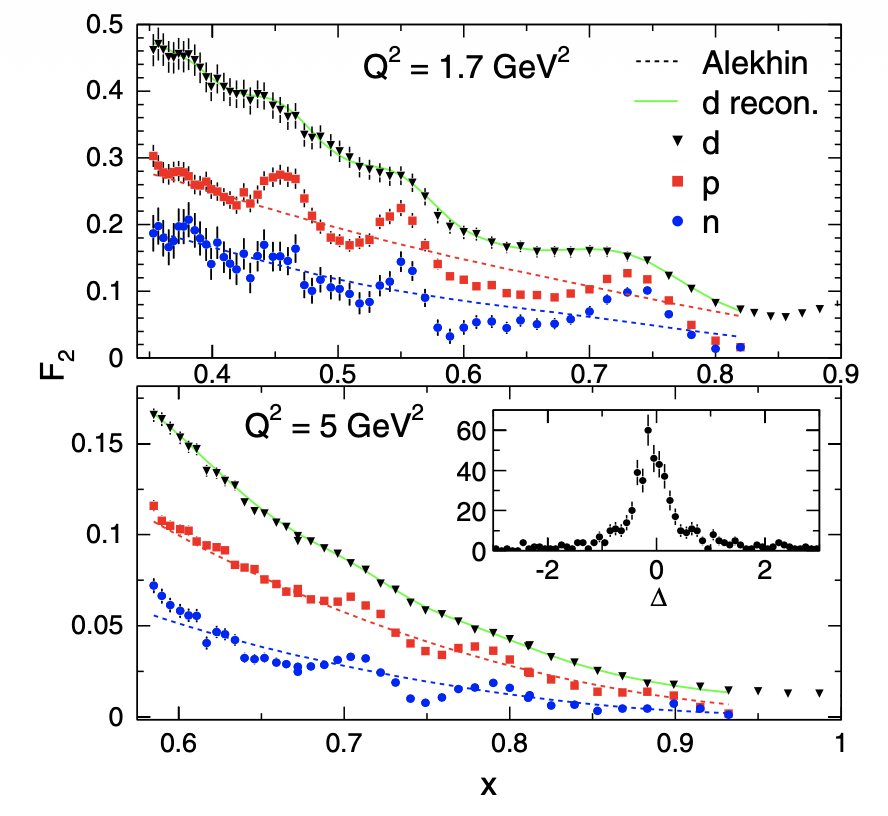}
\vspace*{-0.2cm}
\caption{Proton (blue circles), deuteron (black triangles), and extracted neutron (red squares) $F_2$ structure functions in the resonance region versus $x_B$ compared with leading-twist parametrizations from [\cite{Alekhin:2001cp, Alekhin:2005fp}] (dotted lines). The dependence of the iteration for the neutron extraction on the initial value is illustrated in the inset. Adapted from~[\cite{Malace:2009dg}].}
\label{fig:duality_F2}
\end{figure}

\begin{figure}[t]
\centering
\includegraphics[width=0.54\textwidth]{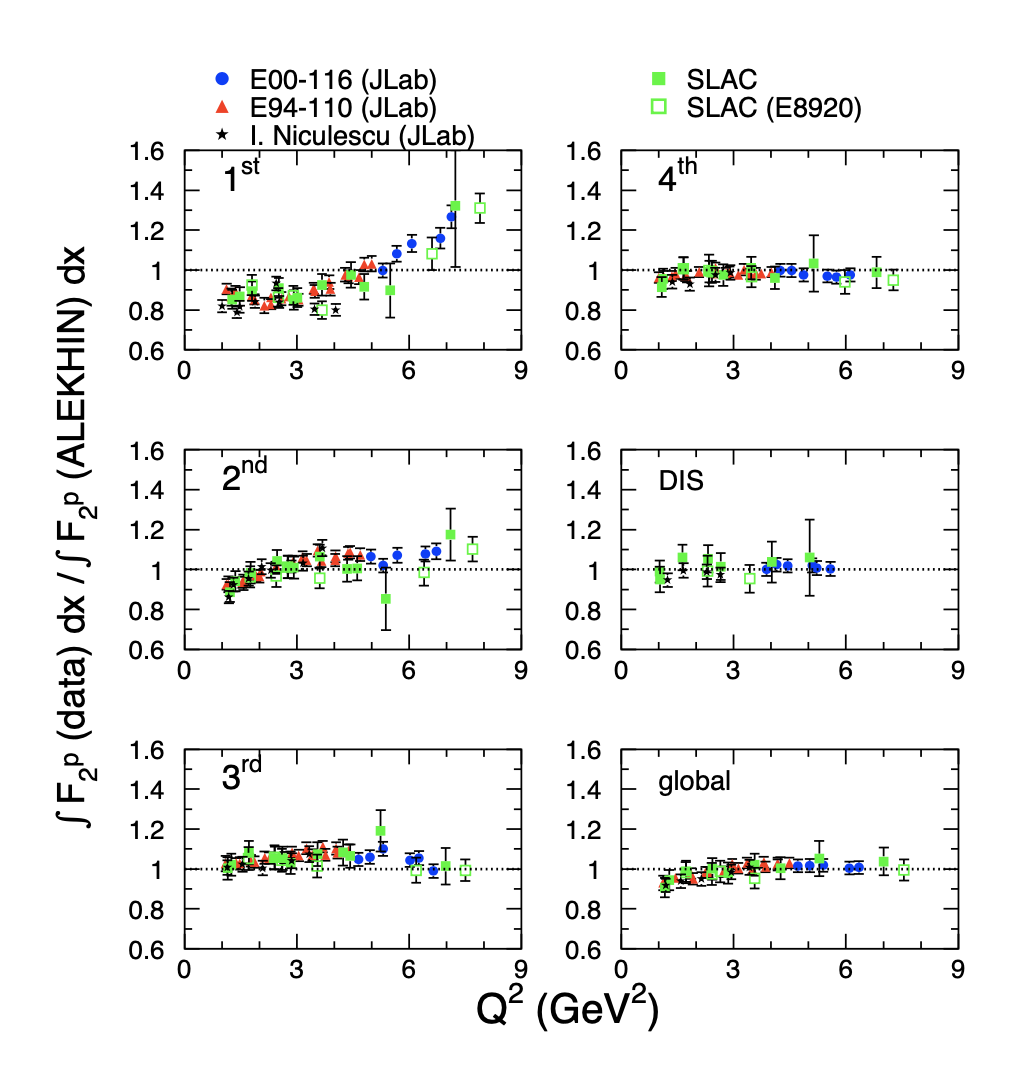}
\vspace*{-0.3cm}
\caption{Ratios of truncated moments of the experimental proton $F_2$ structure function data from Jefferson Lab and SLAC relative to the leading-twist parametrization from [\cite{Alekhin:2001cp, Alekhin:2005fp}] for the 1st, 2nd, 3rd and 4th resonance regions, along with the DIS region, and the global result integrating over all of the measured regions. Adapted from~[\cite{Malace:2009kw}]. \\}
\label{fig:dualratio}
\end{figure}

Quantitative studies compare measured resonance-region structure functions with leading-twist predictions obtained from global PDF analyses after applying perturbative QCD evolution together with TMCs. A comparison of the proton, deuteron, and extracted neutron $F_2$ structure functions from Jefferson Lab data is shown in Fig.~\ref{fig:duality_F2}. The degree of duality can be further quantified through ratios of integrals over the resonance region,
\begin{equation}
R(Q^2)
= \frac{ \displaystyle \int_{\Delta W} dx_B\, F^{\rm exp}(x_B,Q^2)}
       { \displaystyle \int_{\Delta W} dx_B\, F^{\rm LT}(x_B,Q^2)},
\label{eq:duality_ratio}
\end{equation}
or equivalently through truncated moments. Exact duality corresponds to $R=1$, while deviations from unity provide a quantitative measure of the residual higher-twist contributions after averaging over the selected resonance region. This is illustrated in Fig.~\ref{fig:dualratio}, where we show the ratios of the truncated moments to the PDF parametrization of [\cite{Alekhin:2001cp, Alekhin:2005fp}] for the first four prominent resonance regions, along with the DIS and total measured regions (see also [\cite{Malace:2009kw}]).
Among the individual resonance regions, the second and third resonance regions generally exhibit good agreement with leading-twist predictions, while the $\Delta(1232)$ resonance shows the largest deviations. Integrating over the entire resonance region gives the best realization of duality. These observations have practical importance for global QCD analyses, since they suggest the inclusion of resonance-region data in PDF determinations after appropriate treatment of target mass and higher-twist corrections. The increase in kinematic coverage would have the potential for significantly improved constraints on PDFs in the large-$x_B$ region, where conventional DIS measurements become statistically limited.

\begin{figure}[t]
\centering
\includegraphics[width=0.76\textwidth]{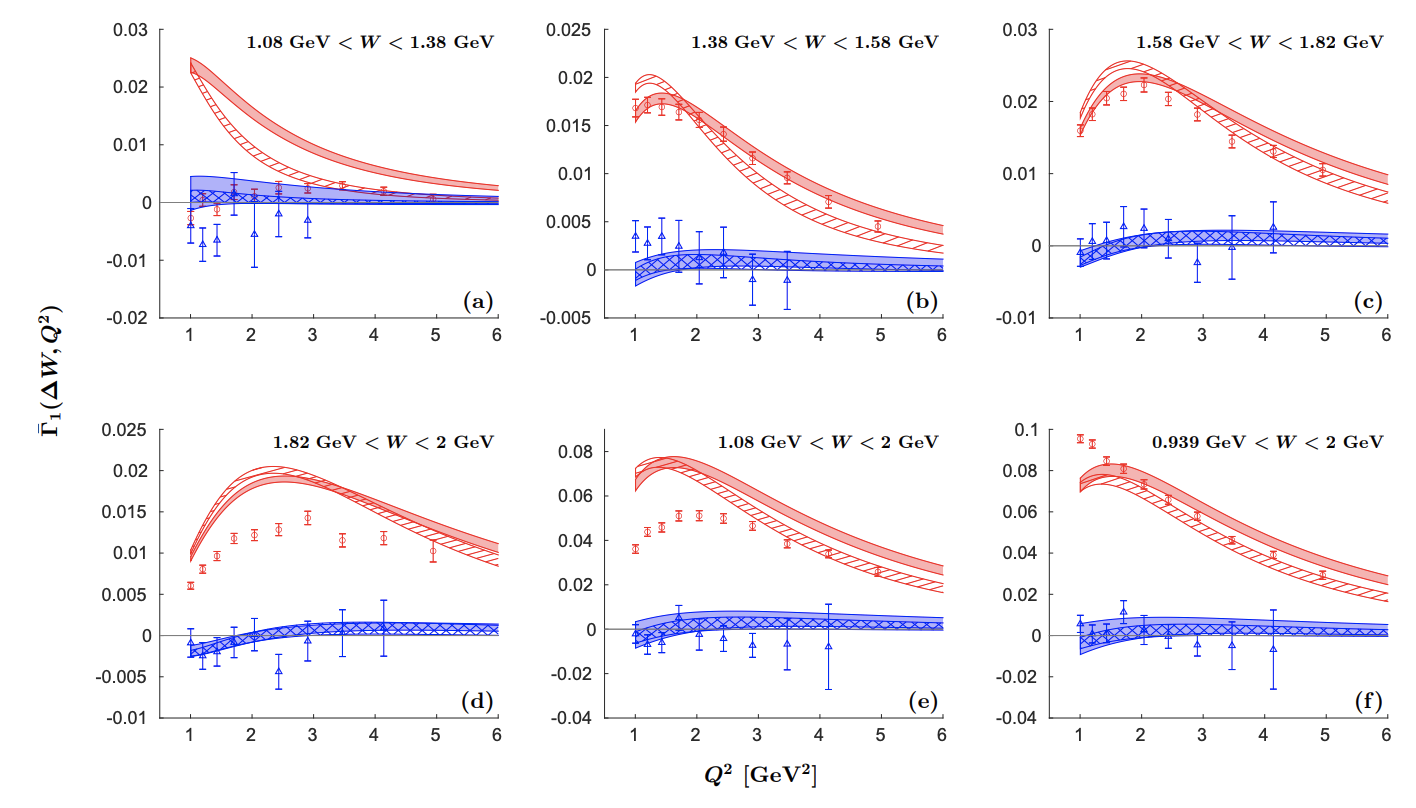}
\vspace*{-0.2cm}
\caption{Comparison of resonance-region contributions to the truncated moment, $\Gamma_1$, of the polarized structure function $g_1$ for the proton (red) and neutron (blue) for various intervals of $W$. The solid bands show the full result from the extrapolated JAM global QCD analysis~[\cite{Cocuzza:2022jye}], including TMC and higher twist contributions, while the hatched bands show only the leading-twist contributions. Figure from~[\cite{Lagerquist:2022tml}].}
\label{fig:g1_duality}
\end{figure}

Quark--hadron duality has also been investigated in polarized DIS through measurements of the spin-dependent structure functions, particularly $g_1$. In contrast to the unpolarized case, the resonance contributions exhibit pronounced sign changes arising from the spin structure of the individual nucleon excitations. For example, the $\Delta(1232)$ resonance contributes negatively at low values of $Q^2$, whereas the scaling function is positive over much of the same kinematic region. Despite these large local differences, the averaged resonance contribution to the $g_1$ integral,
\begin{equation}
{\Gamma}_1(\Delta W,Q^2)
= \int_{x_1(W_1,Q^2)}^{x_2(W_2,Q^2)} dx\, g_1(x,Q^2),
\end{equation}
again approaches the leading-twist prediction as $Q^2$ increases [\cite{Dharmawardane:2006zd, Lagerquist:2022tml}], as Fig.~\ref{fig:g1_duality} illustrates. The onset of duality in polarized structure functions generally occurs at somewhat larger momentum transfers than in $F_2$, reflecting the greater sensitivity of spin observables to higher-twist effects and quark--gluon correlations.

The concept of quark--hadron duality extends naturally to weak interactions, where neutrino scattering provides access to additional structure functions, including the parity-violating structure function $F_3$. Although the available neutrino data are considerably less precise than those from charged-lepton scattering and are generally obtained on nuclear targets, existing analyses~[\cite{Lalakulich:2006yn}] indicate that duality is satisfied at a level comparable to that observed in electromagnetic interactions once nuclear effects are taken into account. The successful description of the transition between the resonance and DIS regions is particularly important for precision neutrino oscillation experiments, where accurate event generators require reliable models spanning the full kinematic range.

Overall, the experimental evidence accumulated over the past decades demonstrates that quark--hadron duality is a robust and widely observed feature of inclusive lepton scattering. Its validity has been established for unpolarized and polarized structure functions, for both proton and neutron targets, and in electromagnetic and weak interactions. Perhaps most significantly, modern high-precision measurements have shown that the effective higher-twist contributions remain surprisingly small after averaging over resonance regions, providing strong support for the OPE-based interpretation of duality. This conclusion has important implications for both the phenomenology of higher twists and the extraction of PDFs from large-$x_B$ data, topics that will be discussed in the following section.

\section{Phenomenology of Higher Twists in DIS}
\label{sec:pheno}

The OPE provides a systematic framework for describing higher-twist contributions to DIS structure functions, but their nonperturbative matrix elements must be determined phenomenologically through comparisons with experimental data. The greatest sensitivity to higher twists is provided by inclusive DIS measurements at moderate $Q^2$ and large $x_B$, where power corrections become comparable with the experimental precision. This section reviews the phenomenological extraction of higher twists from inclusive DIS data and the principal results obtained for unpolarized and polarized structure functions.

\subsection{Extraction of Higher Twists from Experiment}
\label{sec:HTextraction}

As discussed in the previous sections, in phenomenological analyses the structure functions are generally written as the sum of the leading-twist contribution, including TMCs, together with a power-suppressed terms,
\begin{equation}
F_i(x_B,Q^2) 
= F_i^{\rm LT+TMC}(x_B,Q^2) 
+ \frac{H_i(x_B)}{Q^2}
+ {\cal O}\!\left(\frac{1}{Q^4}\right),
\label{eq:HTparam}
\end{equation}
where $i=1,2,L$ and the effective higher-twist function $H(x_B)$ represents the combined contribution of twist-four operators. Some analyses instead adopt a multiplicative parameterization,
\begin{equation}
F_i(x_B,Q^2)
= F_i^{\rm LT+TMC}(x_B,Q^2)
\left[ 1+\frac{C_i(x_B)}{Q^2} \right],
\label{eq:HTmult}
\end{equation}
although in principle both forms should provide comparable descriptions of data if sufficiently flexible parameterizations are employed.

In practice, the extracted higher-twist contributions are not unique, but depend on the theoretical framework adopted in the analysis. In particular, the results are correlated with the choice of leading-twist PDFs, the order of perturbative QCD employed in the Wilson coefficient functions and evolution equations, the prescription used to implement TMCs and nuclear effects (for analyses involving nuclei). Consequently, the fitted higher-twist functions should generally be regarded as effective quantities that absorb all residual contributions not accounted for by the leading-twist description.


The first quantitative extractions of higher twists from high-precision DIS data were performed by~[\cite{Virchaux:1991jc}], who demonstrated the importance of power corrections in describing fixed-target measurements at large $x_B$. Subsequent analyses by~[\cite{Alekhin:2001cp, Alekhin:2005fp}], and the CJ~[\cite{Owens:2012bv, Accardi:2016qay, Cerutti:2025yji, Accardi:2026hdv}] and JAM~[\cite{Sato:2016tuz, Cocuzza:2025qvf, Cocuzza:2026zoy}] collaborations have incorporated increasingly sophisticated treatments of perturbative corrections, target mass effects, and nuclear corrections. Although differences remain among the various analyses, they have established a broadly consistent picture in which higher-twist effects are generally modest over much of the DIS region but become increasingly important at large $x_B$ and moderate $Q^2$, where they are essential for a quantitative description of the data. The phenomenology of higher twists extracted from unpolarized and polarized DIS measurements is reviewed in the following subsections.

\subsection{Higher Twists in Unpolarized DIS}
\label{sec:HTunpol}

The most extensive phenomenological information on higher twists has been obtained for the unpolarized structure function $F_2$. Precise measurements from fixed target experiments span a broad range of $x_B$ and $Q^2$, allowing the logarithmic scaling violations generated by perturbative QCD to be separated from contributions that fall approximately as inverse powers of $Q^2$. One of the first precision extractions was performed by [\cite{Virchaux:1991jc}], who carried out a next-to-leading order (NLO) QCD analysis of high-statistics hydrogen and deuterium $F_2$ data, parametrizing the higher-twist contribution as in Eq.~(\ref{eq:HTmult}). Their results 
provided early quantitative evidence that the fixed-target data could not be described over the full fitted range by perturbative scaling violations and TMCs alone, also demonstrating the correlation between the fitted higher-twist contribution and $\alpha_s$ in a simultaneous determination of both.

\begin{figure}[t]
\centering
\includegraphics[width=0.6\textwidth]{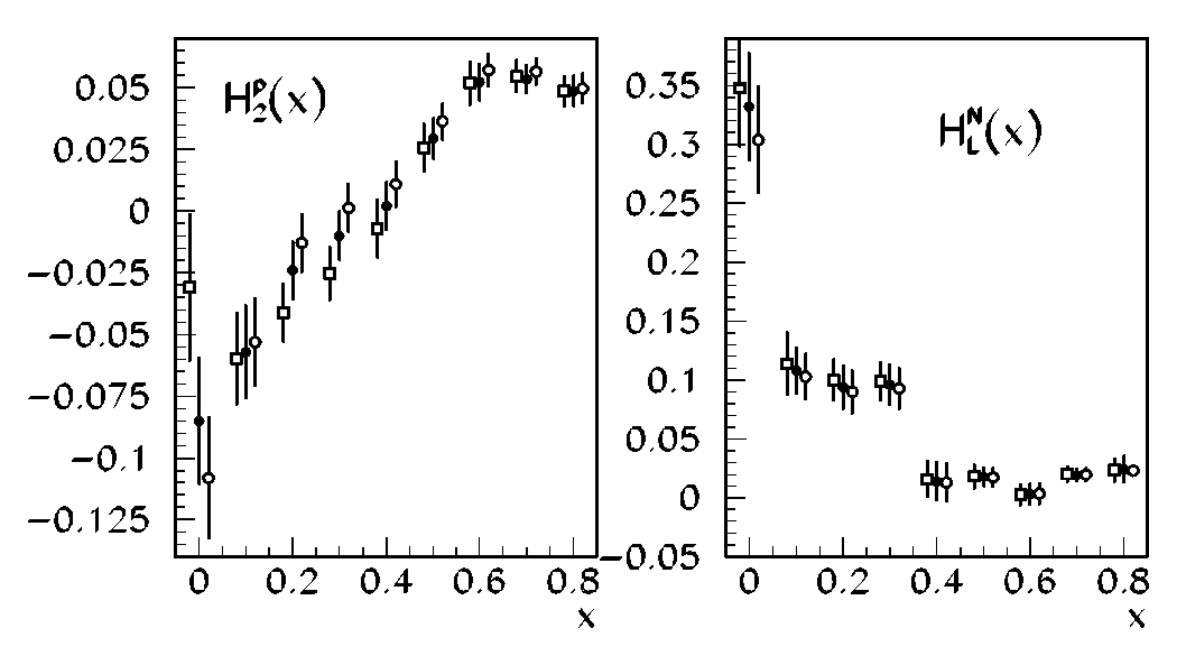}
\caption{Extracted twist-4 contributions to the proton $F_2$ and nucleon $F_L$ structure functions versus $x_B$ for different choices of the renormalization scale: $\mu_R=Q$ (filled circles), $\mu_R=2Q$ (open circles), $\mu_R=Q/2$ (open squares), from [\cite{Alekhin:2001cp}].}
\label{fig:HT_alekhin}
\end{figure}

Later, using broader DIS datasets and more flexible leading-twist and higher-twist parametrizations, [\cite{Alekhin:2001cp, Alekhin:2005fp}]  found that leading-twist QCD supplemented by TMCs alone is insufficient to describe fixed-target measurements, and that phenomenological twist-four contributions are required, particularly at large $x_B$. As illustrated in Fig.~\ref{fig:HT_alekhin}, the extracted higher-twist corrections were found to be generally small over most of the DIS region but to increase rapidly in the valence region, with their quantitative magnitude depending on the perturbative order and theoretical framework adopted in the analysis.

\begin{figure}[b]
\centering
\includegraphics[width=0.38\textwidth]{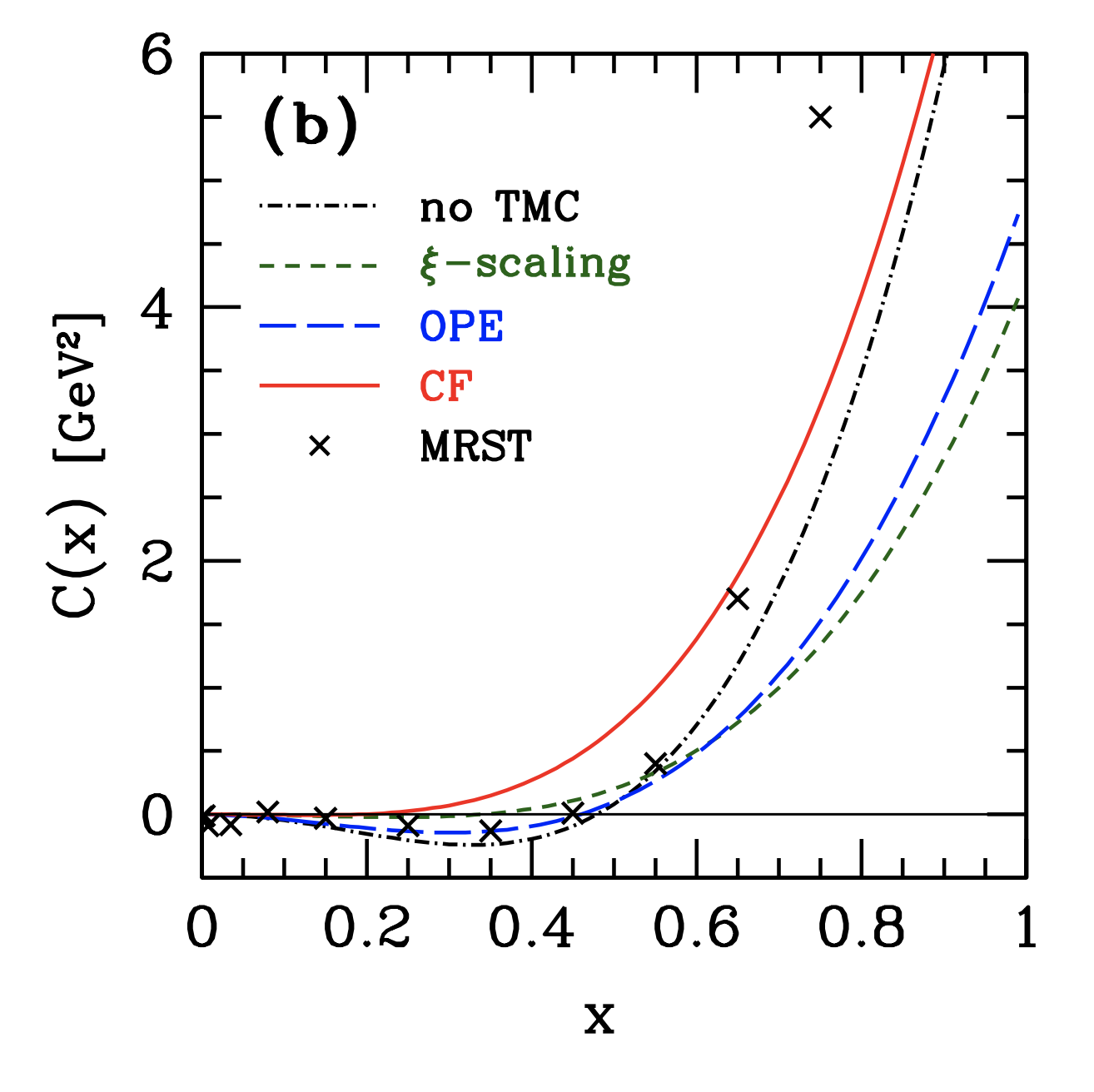}
\caption{Extracted multiplicative higher-twist coefficient $C_2$ for the proton for different TMC prescriptions from the CJ analysis [\cite{Accardi:2009br}], compared with the higher-twist determination from the MRST fit [\cite{Martin:1998np}].}
\label{fig:HT_cj10}
\end{figure}

Modern global QCD analyses have extended the phenomenology to lower values of $W^2$ and larger values of $x_B$ by fitting the higher-twist terms simultaneously with the PDFs and nuclear corrections. The CJ analyses~[\cite{Accardi:2009br, Owens:2012bv, Accardi:2016qay, Cerutti:2025yji}], for example, demonstrated that the inclusion of TMCs, nuclear effects, and phenomenological higher twists allows fixed-target DIS data with substantially relaxed invariant-mass cuts, $W^2 \gtrsim (3-3.5)$~GeV$^2$, to constrain the large-$x$ PDFs. These studies also highlighted correlations between the neutron higher-twist contribution, the deuteron off-shell correction, and the extracted $d/u$ quark PDF ratio. A change in one component can be partially compensated by adjustments in the others while maintaining a comparable description of the measured proton and deuteron cross sections, as Fig.~\ref{fig:HT_cj10} illustrates.

\begin{figure}[t]
\centering
\includegraphics[width=0.65\textwidth]{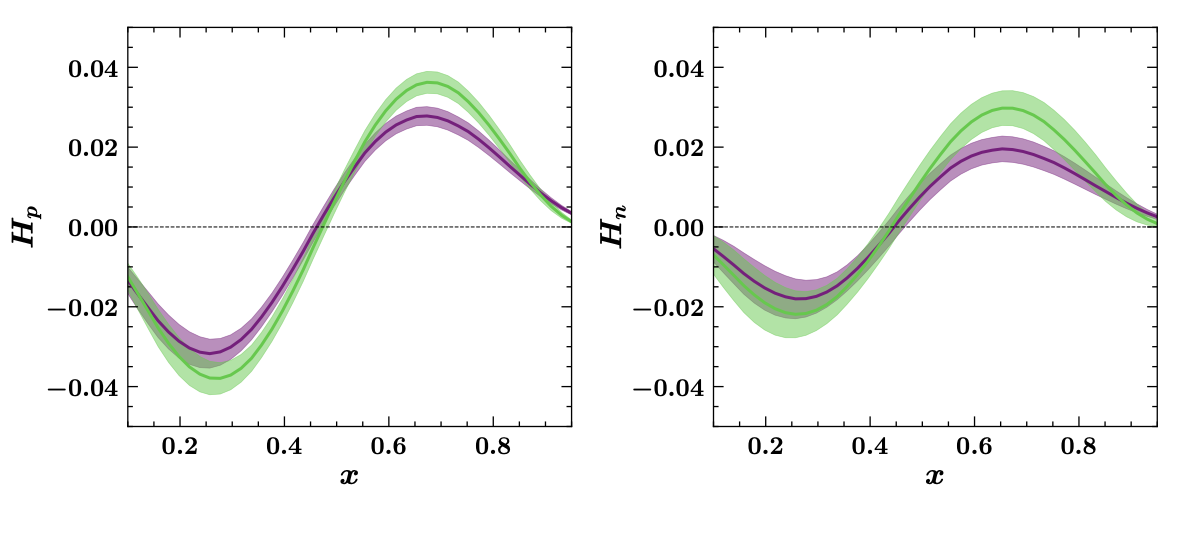}
\caption{Extracted higher-twist coefficient from the CJ25 analysis for the proton and neutron, assuming isospin-dependent additive (green band) or multiplicative (violet band) corrections. Adapted from [\cite{Cerutti:2025yji}].}
\label{fig:HT_cj25}
\end{figure}

A more recent CJ study [\cite{Cerutti:2025yji}] demonstrated that there is no fundamental theoretical preference for additive or multiplicative phenomenological higher-twist parameterizations. Rather, the dominant source of systematic uncertainty arises from imposing isospin-independent higher-twist functions. Such assumptions generate parameterization-dependent distortions of the extracted neutron structure function, $d/u$ ratio, and nucleon off-shell corrections. Allowing independent proton and neutron higher-twist functions largely removes these biases, after which additive and multiplicative parameterizations yield comparable descriptions of the DIS data and consistent physical conclusions. The resulting proton and neutron higher-twist functions from this analysis are shown in Fig.~\ref{fig:HT_cj25}. The higher-twist functions change sign with $x_B$, from negative at $x_B \lesssim 0.4$ to positive in the valence region, and reaching a maximum around $x_B \approx 0.6$--0.7 with magnitudes of order $0.02$--$0.04~\mathrm{GeV}^2$, before decreasing toward zero as $x_B \to 1$. The neutron contribution has a similar $x_B$ dependence, but is generally
smaller than that for the proton, with the additive and multiplicative
parameterizations yielding consistent higher-twist functions within uncertainties.

The most recent JAM analysis [\cite{Cocuzza:2026zoy}] has provided the most comprehensive determination to date of higher-twist effects in unpolarized DIS by combining world proton, deuteron, and $A=3$ data with the latest Jefferson Lab measurements extending to $x_B \approx 0.85$. Within the Bayesian Monte Carlo framework, higher-twist contributions were fitted simultaneously with the leading-twist PDFs and nucleon off-shell corrections, allowing a systematic study of the correlations among these effects. The analysis demonstrated that inclusive DIS data can be consistently described down to $W^2=3.5~{\rm GeV}^2$ and $Q^2=m_c^2$, thereby substantially extending the kinematic range available for constraining the large-$x$ PDFs. The baseline fit employed an additive higher-twist parameterization with independent proton and neutron higher-twist functions, together with TMCs implemented within the collinear factorization framework [\cite{Moffat:2019jto}].

\begin{figure}[b]
\centering
\includegraphics[width=0.75\textwidth]{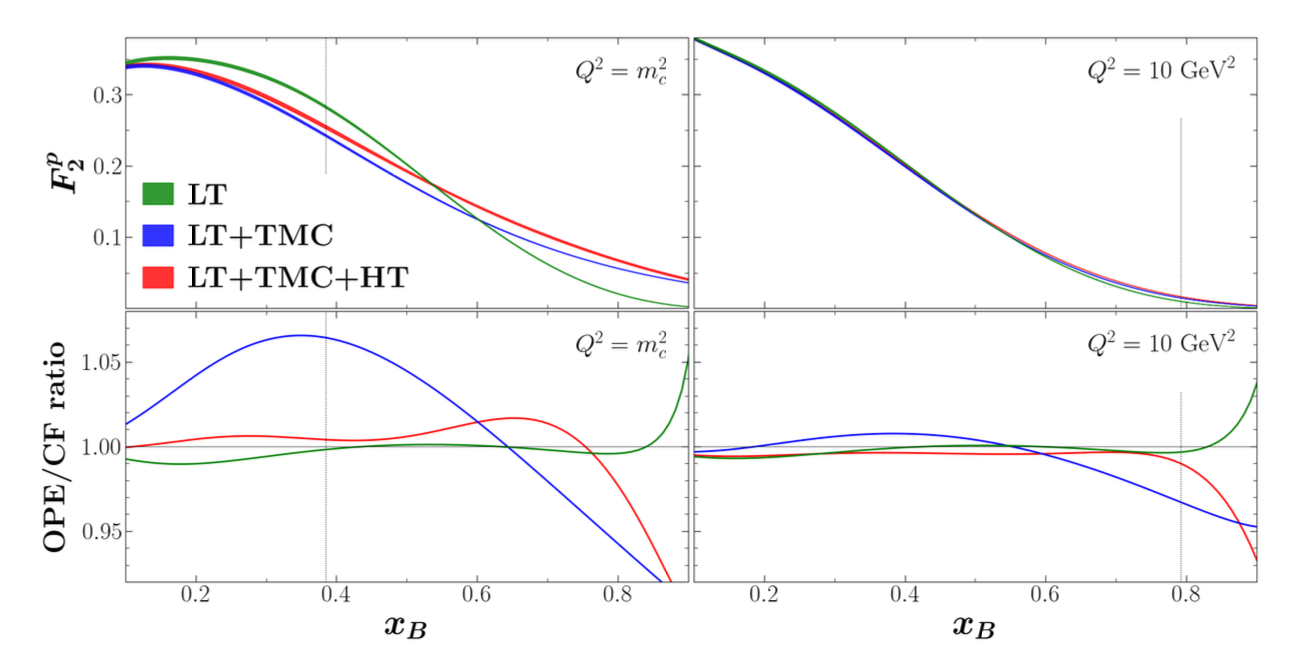}
\caption{{\it [Top row]} Proton structure function $F_2^p$ as a function of $x_B$ for the JAM fit [\cite{Cocuzza:2026zoy}] at $Q^2 = m_c^2$ {\it (left column)} and $Q^2=10$~GeV$^2$ {\it (right column)}. The full JAM results (red 68\% CI bands) are compared to the same fits with only the leading twist (LT) contribution (green bands) and with LT and TMCs included (blue bands). {\it [Bottom row]} Ratio of the mean of the fit with OPE TMCs to the mean of the JAM fit with the collinear factorization TMCs. The vertical lines indicate the values of $x_B$ corresponding to the resonance region cut of $W^2=3.5$~GeV$^2$. Adapted from [\cite{Cocuzza:2026zoy}].}
\label{fig:HT_jamF2}
\end{figure}

\begin{figure}[t]
\centering
\includegraphics[width=0.44\textwidth]{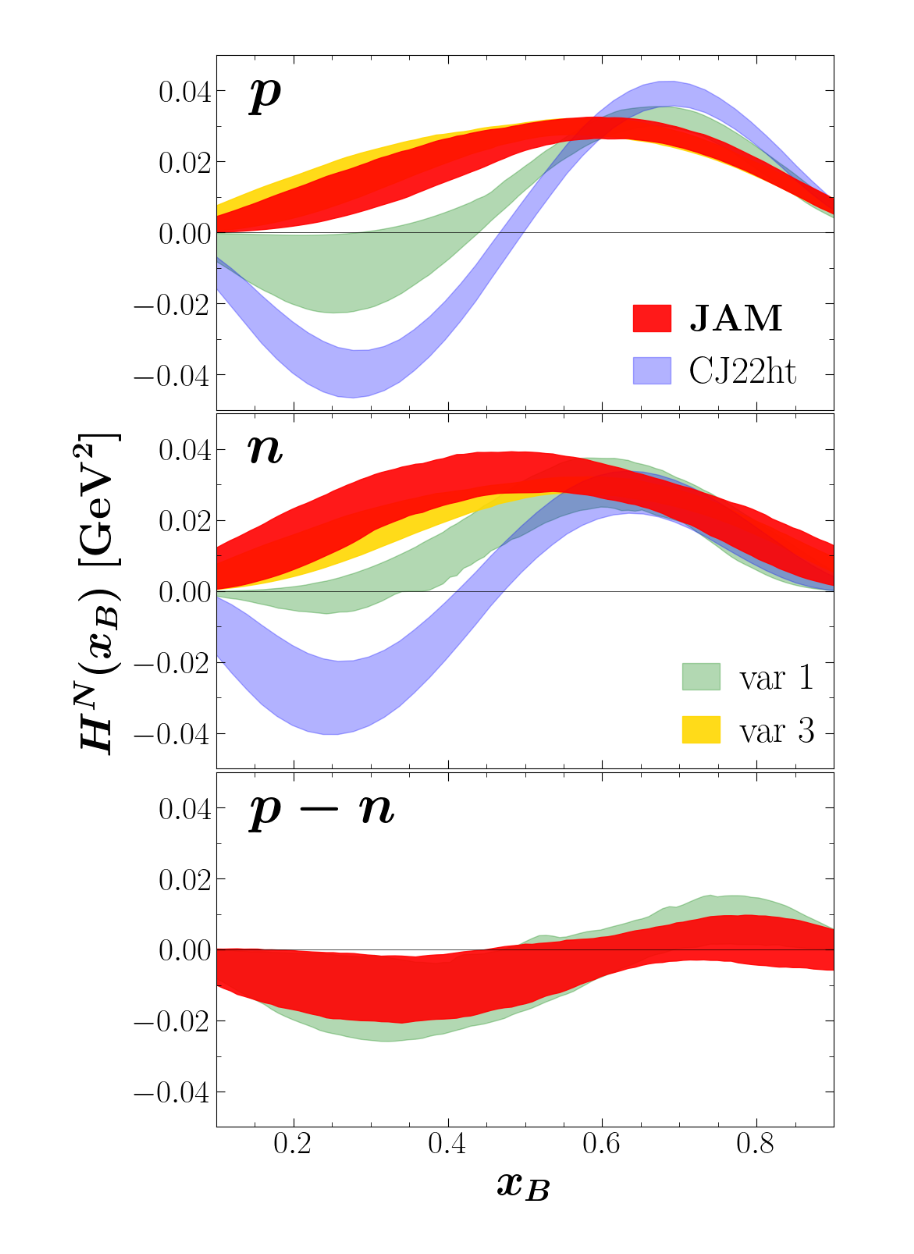}
\caption{Additive higher twist function $H^N(x_B)$ for the proton $p$ (top), neutron $n$ (middle), and $p-n$ difference (bottom) for JAM (red bands), compared with the fits for ``var~1'' (using OPE TMCs, green bands) and ``var~3'' (using isospin symmetric higher twists, yellow bands), and the CJ22ht results~[\cite{Cerutti:2025yji}] (blue bands). Adapted from [\cite{Cocuzza:2026zoy}].}
\label{fig:HT_jamH}
\end{figure}

The decomposition of the proton and neutron structure functions shown in Fig.~\ref{fig:HT_jamF2} at the input scale $Q^2=m_c^2$ and at $Q^2=10$~GeV$^2$ illustrates the relative importance of the leading-twist, TMC, and higher-twist contributions over the kinematic range of the fit. At low and intermediate $x_B$, the leading-twist contribution already provides a good description of the structure functions, with both TMCs and higher twists becoming negligible as $Q^2$ increases. In the valence region, however, finite-$Q^2$ effects become increasingly important: TMCs account for a substantial fraction of the observed corrections, while the remaining difference is described by a positive higher-twist contribution that decreases rapidly with increasing $Q^2$. The extracted higher-twist functions shown in Fig.~\ref{fig:HT_jamH} are positive over the fitted $x_B$ range for both the proton and neutron, with only a small proton--neutron difference that is consistent with zero within uncertainties. The analysis also demonstrated the strong correlation between TMCs and higher twists, with the use of OPE TMCs yielding substantially smaller, and in some regions negative, higher-twist functions while leaving the full fitted structure functions essentially unchanged. Finally, the JAM study found that additive and multiplicative higher-twist parameterizations provide equally good descriptions of the data, indicating that the present DIS measurements do not discriminate between these parametrizations when sufficient flexibility is allowed in the fit.

\subsection{Higher Twists in Polarized DIS}
\label{sec:HTpol}

Polarized DIS provides unique information on higher twists through moments of the spin-dependent structure functions. In particular, the lowest moment of $g_1$ allows the leading power correction to be related to reduced matrix elements with a direct interpretation in terms of spin-dependent quark--gluon correlations. The lowest moment is defined by
\begin{equation}
\Gamma_1(Q^2)
= \int_0^1 dx_B\, g_1(x_B,Q^2),
\label{eq:Gamma1_pol}
\end{equation}
which includes the elastic contribution at $x_B=1$. Within the OPE, the moment can be expanded as
\begin{equation}
\Gamma_1(Q^2)
= \mu_2(Q^2) + \frac{\mu_4(Q^2)}{Q^2} + \frac{\mu_6(Q^2)}{Q^4} + \cdots,
\label{eq:Gamma1_OPE}
\end{equation}
where $\mu_2$ denotes the leading-twist contribution, including its perturbative QCD evolution, and the coefficients $\mu_{\tau}$ contain matrix elements of operators of increasing twist. At leading order in $\alpha_s$, the first moment is related to the helicity-dependent quark distributions,
\begin{equation}
\mu_2(Q^2)
= \frac{1}{2} \sum_q e_q^2 \int_0^1 dx_B\, \Delta q^+(x_B,Q^2),
\label{eq:Gamma1_LT}
\end{equation}
up to the perturbatively calculable Wilson coefficients, where $\Delta q^+ = \Delta q + \Delta \bar q$. Equivalently, it may be expressed in terms of the axial charges of the nucleon [\cite{Anselmino:1994gn, Lampe:1998eu}],
\begin{equation}
\mu_2^{p(n)} = \pm\frac{1}{12} g_A + \frac{1}{36} a_8 + \frac{1}{9} \Delta\Sigma,
\end{equation}
where 
$g_A = \Delta u^+ - \Delta d^+$ is triplet axial charge, 
$a_8 = \Delta u^+ + \Delta d^+ - 2\Delta s^+$ is the octet charge,
and 
$\Delta\Sigma = \Delta u^+ + \Delta d^+ + \Delta s^+$ is the singlet axial charge, corresponding to the total helicity of the nucleon carried by quarks and antiquarks.

In the OPE the leading $1/Q^2$ correction to the first moment can be written as
\begin{equation}
\mu_4(Q^2)
= \frac{M^2}{9} \left[ a_2(Q^2) + 4 d_2(Q^2) + 4 f_2(Q^2) \right].
\label{eq:mu4_pol}
\end{equation}
The quantity $a_2$ is a twist-two matrix element associated with TMCs, while the matrix element $d_2$ has twist three and is obtained from the $x_B^2$-weighted combination of the measured $g_1$ and $g_2$ structure functions,
\begin{equation}
d_2(Q^2)
= \int_0^1 dx_B\, x_B^2 \left[ 2 g_1(x_B,Q^2) + 3 g_2(x_B,Q^2) \right].
\label{eq:d2_pol}
\end{equation}
The leading-twist Wandzura--Wilczek contribution to $g_2$,
\begin{equation}
g_2^{\rm WW}(x_B,Q^2)
= -g_1(x_B,Q^2) + \int_{x_B}^{1} \frac{dy}{y}\, g_1(y,Q^2),
\label{eq:WW}
\end{equation}
cancels in this combination, so that a nonzero $d_2$ measures genuine twist-three quark--gluon correlations. Because of the $x_B^2$ weighting, the integral is particularly sensitive to the large-$x_B$ and resonance regions. The remaining matrix element $f_2$ is associated with a twist-four quark--gluon operator defined by
\begin{equation}
f_2(Q^2)\,M^2 S^\mu
= \frac12 \sum_q e_q^2 
\left\langle P,S \left| g\,\bar q\, \widetilde G^{\mu\nu} \gamma_\nu q
\right| P,S \right\rangle ,
\label{eq:f2_operator}
\end{equation}
where $\widetilde G^{\mu\nu}$ is the dual gluon field-strength tensor. Unlike $a_2$, which is kinematic in origin, both $d_2$ and $f_2$ contain information on dynamical correlations between the struck quark and the color fields generated by the remaining partons.

The extracted $d_2$ and $f_2$ matrix elements can also be expressed in terms of the color electric and magnetic polarizabilities, 
$\chi_E = \frac13 (4 d_2 + 2 f_2)$ and 
$\chi_B = \frac13 (4 d_2 - f_2)$, which characterize the response of the color fields inside the nucleon to its spin [\cite{Ji:1994br}]
(see also [\cite{Burkardt:2008ps}]). 
%
%
The possibility of extracting these matrix elements from polarized moments was discussed by [\cite{Ji:1994br}], who suggested that resonance-region data could be used to determine higher-twist matrix elements once the leading-twist contribution is specified. Subsequently [\cite{Ji:1997gs}] applied this method to the available proton and neutron spin-structure data to extract values of the $f_2$ matrix element and the polarizabilities $\chi_E$ and $\chi_B$.

\begin{figure}[t]
\centering
\includegraphics[width=0.43\textwidth]{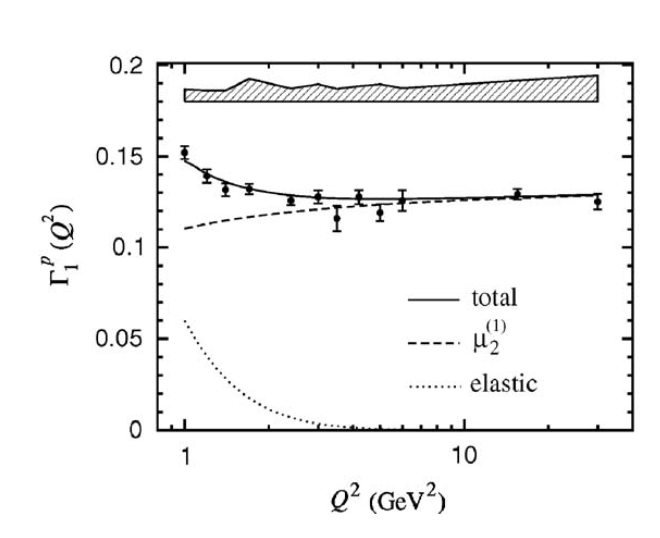}
\includegraphics[width=0.42\textwidth]{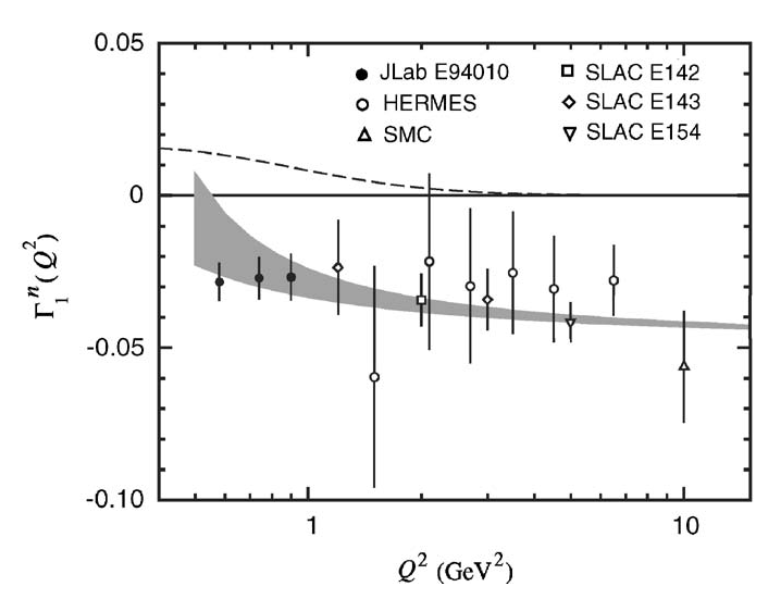}
\caption{(Left panel) Lowest moment of the proton $g_1^p$ structure function. The points are from a reanalysis of world data by [\cite{Osipenko:2004xg, Osipenko:2005nx}]; the error bars give statistical uncertainties only, while the systematic and low-$x_B$ extrapolation errors are given by the shaded band. (Right panel) Lowest moment of the neutron $g_1^n$ structure function [\cite{Meziani:2004ne}]. The error bars are a quadratic sum of statistical and systematic errors. The shaded band represents the uncertainty on the leading-twist contribution due to $\alpha_s$, and the dashed curve indicates the elastic contribution. Adapted from [\cite{Melnitchouk:2005zr}].}
\label{fig:HT_Gam1}
\end{figure}

More comprehensive extractions became possible with the availability of precise Jefferson Lab data in the resonance region. A reanalysis of the world proton data, including measurements over $1 < Q^2 < 30~{\rm GeV}^2$, found that the first moment is effectively dominated by its leading-twist contribution for $Q^2\gtrsim2$--$3~{\rm GeV}^2$, while a discernible power correction appears below this range [\cite{Osipenko:2004xg, Osipenko:2005nx}]. For the neutron, the inclusion of Jefferson Lab resonance-region data allowed the first moment to be analyzed down to $Q^2\approx0.5~{\rm GeV}^2$ [\cite{Meziani:2004ne}], finding that the resulting twist-four matrix element was consistent with zero within the uncertainty. The $\Gamma_1^p$ and $\Gamma_1^n$ data from these analyses are shown in Fig.~\ref{fig:HT_Gam1}. Combining the extracted $f_2$ values with the available measurements of $d_2$ gave $\chi_E^p = 0.026(28)$ and $\chi_B^p = -0.013(14)$ for the proton, and $\chi_E^n = 0.033(29)$ and $\chi_B^n =-0.001(16)$ for the neutron electric and magnetic responses.

Turning now to the polarized structure functions themselves, early $x_B$-dependent extractions of higher-twist contributions were performed by the LSS group [\cite{Leader:2005ci, Leader:2010rb}], who parametrized the structure function as the sum of the leading-twist contribution, including TMCs, and an additive higher-twist term,
\begin{equation}
g_1(x_B,Q^2) = g_1^{\rm LT+TMC}(x_B,Q^2) + \frac{h(x_B)}{Q^2},
\label{eq:g1HT_add}
\end{equation}
and extracted the function $h(x_B)$ simultaneously with the polarized PDFs from global fits to polarized DIS data. The extracted higher-twist corrections were found to be most significant in the valence region, particularly for the neutron, while remaining relatively small at low and intermediate $x_B$. The inclusion of these corrections was shown to improve the description of Jefferson Lab data at moderate $Q^2$.

An alternative analysis by Bl\"umlein and B\"ottcher employed a multiplicative parameterization [\cite{Blumlein:2010rn}],
\begin{equation}
g_1(x_B,Q^2)
= g_1^{\rm LT}(x_B,Q^2) \left[ 1+\frac{C(x_B)}{Q^2} \right],
\label{eq:g1HT_mult}
\end{equation}
and concluded that the available polarized DIS data were generally consistent with small higher-twist corrections. This result prompted a detailed comparison by LSS [\cite{Leader:2010rb}], who pointed out that the additive and multiplicative forms are not equivalent and therefore the extracted functions $h(x_B)$ and $C(x_B)$ cannot be directly compared. They argued that the additive form follows more naturally from the OPE, in which the leading- and higher-twist contributions appear as separate terms in the $1/Q^2$ expansion of the structure functions. Despite the differences in methodology, both analyses concluded that higher-twist effects are primarily confined to the large-$x_B$ region and become essential for describing the precision Jefferson Lab measurements at moderate $Q^2$.

More recently, the JAM collaboration performed a comprehensive analysis [\cite{Cocuzza:2025qvf}] of the world's inclusive and semi-inclusive DIS data, along with inclusive jet and $W/Z$ boson production data from polarized $pp$ collisions. Unlike earlier analyses that extracted only an effective higher-twist correction to $g_1$, the JAM analysis parametrized both the twist-four contribution to $g_1$ and the twist-three contribution to $g_2$ simultaneously, with the total polarized structure functions computed as
\begin{equation}
g_i(x_B,Q^2)
= g_i^{\rm LT+TMC}(x_B,Q^2) + g_i^{\rm HT}(x_B,Q^2),
\qquad i=1,2,
\label{eq:JAM_total}
\end{equation}
where $g_i^{\rm LT+TMC}$ denotes the leading-twist structure function including TMCs. While earlier JAM studies~[\cite{Sato:2016tuz}] of DIS data used the OPE framework for the TMCs to the polarized structure functions~[\cite{Wandzura:1977qf, Piccione:1997zh, Blumlein:1998nv}], for consistency when analyzing also non-DIS data the new JAM analysis employed the collinear factorization framework for the TMCs. As in the previous LSS analyses, the new JAM analysis introduced phenomenological higher-twist contributions through additive corrections,
\begin{equation}
g_1^{\rm HT}(x_B,Q^2)
=
\frac{c_1^{\rm HT}(x_B)}{Q^2},
\qquad
g_2^{\rm HT}(x_B,Q^2)
=
c_2^{\rm HT}(x_B),
\label{eq:JAM_HT}
\end{equation}
where the functions $c_1^{\rm HT}(x_B)$ and $c_2^{\rm HT}(x_B)$ were determined phenomenologically from the data. For $g_2$, the leading-twist contribution is given by the Wandzura--Wilczek term,
\begin{equation}
g_2^{\rm LT}(x_B,Q^2) = g_2^{\rm WW}(x_B,Q^2),
\label{eq:JAM_g2LT}
\end{equation}
as in Eq.~(\ref{eq:WW}), so that $g_2^{\rm HT}$ represents the genuine twist-three contribution, while $g_1^{\rm HT}$ corresponds to the leading dynamical correction suppressed by $1/Q^2$.

\begin{figure}[t]
\centering
\includegraphics[width=0.33\textwidth]{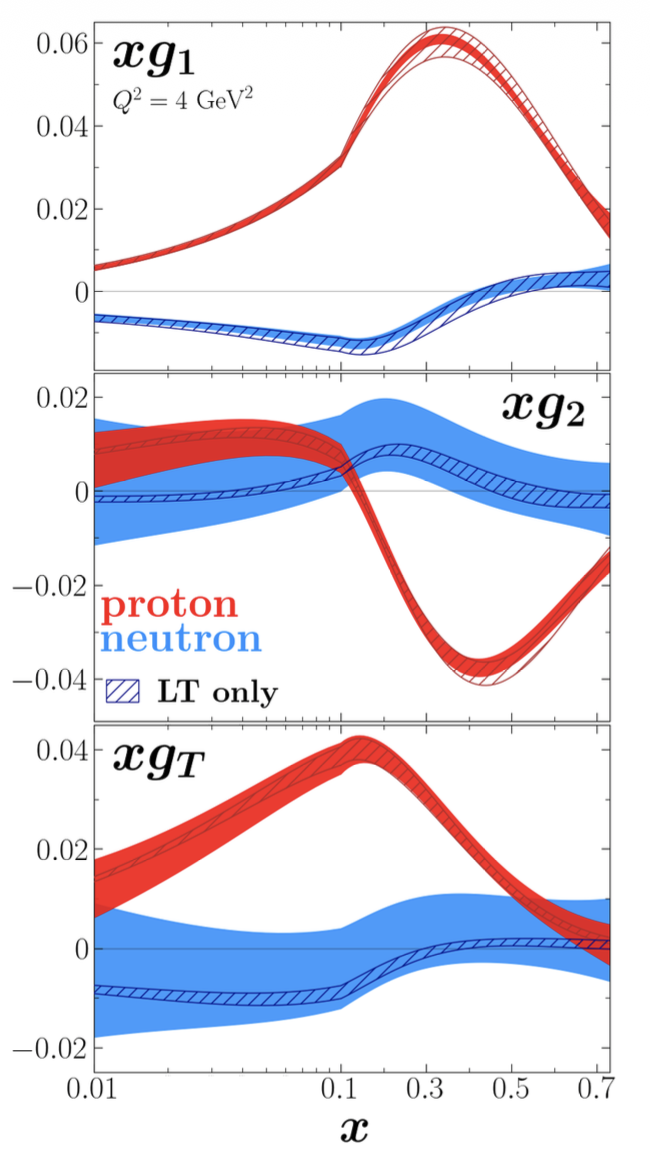}
\vspace*{-0.2cm}
\caption{Polarized structure functions $x_B g_1$ (top panel), $x_B g_2$ (middle panel), and $x_B g_T$ (bottom panel) evaluated at $Q^2=4$~GeV$^2$, for the proton (red) and neutron (blue), with the full JAM results (solid bands) compared with those at LT (hatched bands). Adapted from [\cite{Cocuzza:2025qvf}] (note that $x\equiv x_B$ in this figure).}
\label{fig:jam_g1g2}
\end{figure}

\begin{figure}[h]
\centering
\includegraphics[width=0.62\textwidth]{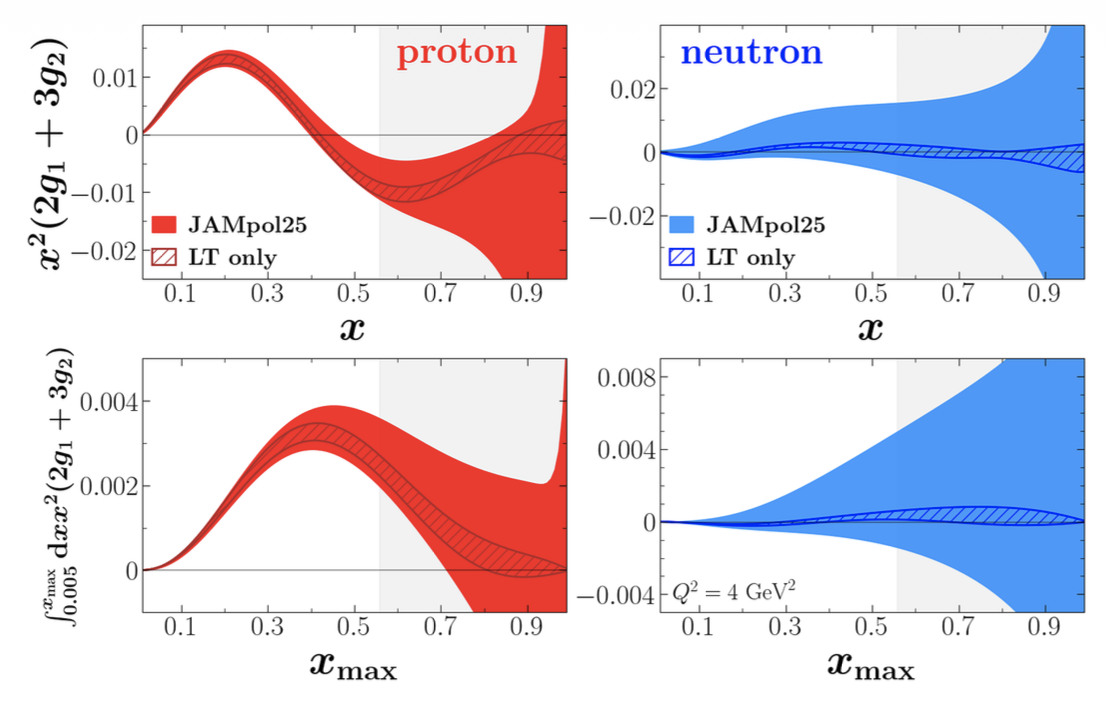}
\vspace*{-0.2cm}
\caption{(Top row) Integrand of the $d_2$ moment for the proton (left) and neutron (right) at $Q^2=4$~GeV$^2$ for the full ``JAMpol25'' fit (solid bands) and the LT approximation (no HT correction or TMC, hatched bands). (Bottom row) Truncated integral as a function of $x_{\rm max}$ starting from $x_B = 0.005$ for the proton (left) and neutron (right). The region of extrapolation into the low-$W$ region is indicated by the gray shaded area. Adapted from [\cite{Cocuzza:2025qvf}] (note that $x\equiv x_B$ in this figure).}
\label{fig:jam_d2}
\end{figure}

The extracted polarized structure functions from the JAM analysis are shown in Fig.~\ref{fig:jam_g1g2} at $Q^2=4~{\rm GeV}^2$, comparing the full fit, including TMCs and higher-twist contributions, with the leading-twist approximation. For both the proton and neutron, the $g_1$ structure functions are found to be remarkably stable with respect to the inclusion of higher-twist effects, with only small modifications to the central values and uncertainties. This indicates that the existing polarized DIS data provide sufficient constraints on the twist-four contribution to $g_1$, which remains relatively modest even in the valence region. The extracted higher-twist correction is slightly negative for the proton and positive for the neutron at large $x_B$, leading to a partial cancellation in the Bjorken sum rule.

The impact of higher twists is more pronounced for the $g_2$ structure function. While the proton $g_2$ is only weakly affected, the inclusion of the twist-three contribution significantly increases the uncertainties on the neutron $g_2$, reflecting the limited experimental constraints from transversely polarized double spin asymmetry measurements on deuteron and $^3$He targets. These effects propagate to the transverse spin structure function, $g_T=g_1+g_2$, for which the proton results remain largely unchanged, whereas the neutron distribution exhibits substantially larger uncertainties once higher-twist effects are included. In particular, a leading-twist analysis alone would suggest a negative neutron $g_T$ for $x_B\lesssim0.2$, while the full JAM analysis shows that $g_T$ remains consistent with zero over the entire measured $x_B$ range once the higher-twist uncertainties are taken into account.

Further insight into the role of higher twists is provided by the truncated $d_2$ moments shown in Fig.~\ref{fig:jam_d2}. The upper panels display the integrands of the $d_2$ matrix element for the proton and neutron, comparing the full JAM analysis with the leading-twist approximation. While the neutron integrand remains consistent with zero over the measured $x_B$ range, the proton integrand exhibits noticeable higher-twist and TMC effects, particularly at large $x_B$. The lower panels show the corresponding truncated moments as functions of the upper integration limit, $x_{\rm max}$, with the lower limit fixed at $x_B=0.005$. For the neutron, the truncated moment is consistent with zero independent of whether higher-twist effects are included. For the proton, the leading-twist approximation yields a positive truncated moment, whereas the inclusion of TMCs and higher twists produces a result that is compatible with zero within the rapidly increasing uncertainties associated with the extrapolation into the unmeasured high-$x_B$ (low-$W$) region. The analysis concludes that the currently available data do not provide statistically significant evidence for a nonzero $d_2$ matrix element, but also demonstrate that future Jefferson Lab 12~GeV measurements extending to higher $x_B$ will be essential for reducing the extrapolation uncertainties and establishing the magnitude of the twist-three quark--gluon correlations.

\section{Higher Twists in Other Processes}
\label{sec:beyond}

Although inclusive DIS has provided the principal laboratory for investigating higher-twist effects, power-suppressed contributions are a generic feature of QCD factorization and arise in a broad class of hard scattering processes. Beyond inclusive reactions, higher twists encode a rich variety of multiparton correlations that become accessible through additional kinematic variables, and exclusive final states. These observables provide complementary information on the nonperturbative structure of hadrons and probe correlation functions that cannot be accessed in inclusive measurements alone.

The theoretical description of higher twists beyond inclusive DIS is considerably richer than for inclusive structure functions. Depending on the process, power corrections involve twist-3 and twist-4 parton distribution and fragmentation functions, TMD distributions, GPDs, and multiparton correlation functions containing explicit gluon fields. In many cases these quantities generate observables that vanish at leading twist, making them particularly sensitive probes of $qg$ correlations and color dynamics inside hadrons. In this section we review several important areas in which higher-twist effects play a central role, beginning with SIDIS, where higher-twist PDFs and fragmentation functions generate a variety of spin-dependent and azimuthal asymmetries. We then discuss higher-twist effects in TMD factorization and in exclusive processes.

\subsection{Semi-Inclusive Deep-Inelastic Scattering}

The SIDIS process,
\begin{equation}
\ell(k) + N(P) \to \ell'(k') + h(p_h) + X,
\end{equation}
extends the physics of inclusive DIS by detecting a hadron $h$ in the final state in coincidence with the scattered lepton. In addition to the usual DIS variables $x_B$, $Q^2$, and $y$, SIDIS introduces the hadron momentum fraction
\begin{equation}
z_h = \frac{P \cdot p_h}{P \cdot q},
\end{equation}
together with the transverse momentum of the produced hadron, $p_{h T}$. These additional degrees of freedom make SIDIS a particularly powerful tool for investigating the multidimensional structure of hadrons and the dynamics of parton fragmentation. The invariant mass squared of the unobserved hadronic system in SIDIS, $W_{\rm SIDIS}^2 \equiv (P+q-p_h)^2$, is given by~[\cite{Whitehill:2022mpq}]
\begin{equation}
\begin{aligned}
W_{\rm SIDIS}^2
=&
M^2 + m_h^2 + Q^2\,\frac{(1-x_B-z_h)}{x_B}
+ 2 \frac{z_h Q^2}{\rho^2-1} 
\left[ \rho \sqrt{1-(\rho^2-1)\frac{m_{hT}^2}{z_h^2\, Q^2}} - 1 \right],
\end{aligned}
\label{eq:WSIDIS_exact}
\end{equation}
where $m_{hT}^2 \equiv m_h^2 + p_{hT}^2$ is the transverse mass squared of the hadron $h$ of mass $m_h$, and $\rho$ is given in Eq.~(\ref{eq:rho2}). In the Bjorken limit ($M/Q$, $m_h/Q$, $p_{hT}/Q$ $\ll 1$), Eq.~(\ref{eq:WSIDIS_exact}) reduces to the simple expression
\begin{equation}
W_{\rm SIDIS}^2
\approx  M^2 + Q^2 \left(\frac{1}{x_B}-1\right)(1-z_h).
\label{eq:WSIDIS_approx}
\end{equation}
Large values of $W_{\rm SIDIS}$ correspond to the current-fragmentation region, while small values signal the onset of resonance production and exclusive channels, where power corrections and possible violations of leading-twist collinear factorization become increasingly important.

\begin{figure}[t]
\centering
\includegraphics[width=0.4\textwidth]{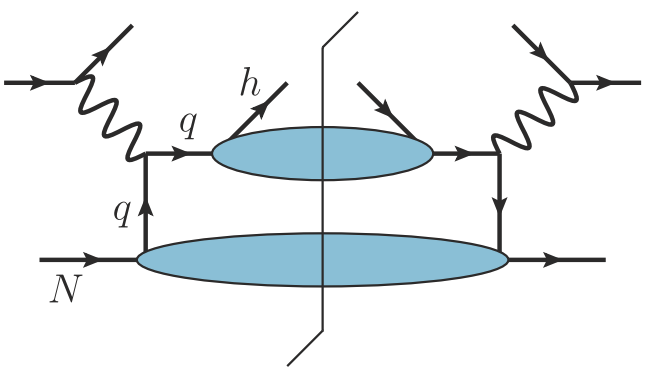}
\caption{Parton-model representation of the leading-twist SIDIS cross section for scattering a lepton from a nucleon $N$ producing a hadron $h$ in the final state. From [\cite{Metz:2016swz}].}
\label{fig:sidis_diag}
\end{figure}

At sufficiently large $Q^2$ the $p_{h T}$-integrated SIDIS cross section factorizes into perturbatively calculable hard-scattering coefficients, PDFs, and fragmentation functions describing the hadronization of the scattered parton into the hadron $h$~[\cite{Collins:1989gx, Collins:2011zzd, Collins:1981uw, Bacchetta:2006tn}]. In collinear factorization, the leading-power contribution may be written schematically as~[\cite{Metz:2016swz}]
\begin{equation}
d\sigma
= \sum_{i} H_{i}(Q^2,\mu) \otimes f_i(x,\mu) \otimes D_i^h(z,\mu)
+ {\cal O}\!\left(\frac{1}{Q}\right),
\label{eq:SIDISfactorization}
\end{equation}
where $H_{i}$ denotes the perturbative hard-scattering kernel, $f_i$ is the PDF, $D_i^h$ is the parton $i$ $\to$ hadron $h$ fragmentation function, and $\mu$ is the factorization scale. This process is illustrated in Fig.~\ref{fig:sidis_diag}. Corrections suppressed by powers of $1/Q$ arise from higher-twist operators and from kinematic effects associated with intrinsic transverse momentum.

The twist expansion of the SIDIS structure functions has the general form
\begin{equation}
F = F^{(\tau=2)} + \frac{F^{(\tau=3)}}{Q} + \frac{F^{(\tau=4)}}{Q^2} + \cdots,
\label{eq:SIDIStwist}
\end{equation}
where the first power correction appears already at order $1/Q$ through twist-3 contributions. This differs from inclusive DIS, for which the leading power corrections enter at order $1/Q^2$. The appearance of twist-3 terms reflects the richer spin and momentum structure of SIDIS, where interference between amplitudes with different orbital angular momentum and quark--gluon correlations gives rise to observables that are absent in inclusive scattering.

At twist three, the SIDIS cross section receives contributions from both distribution and fragmentation correlators. These involve matrix elements containing explicit gluon fields, which describe coherent quark--gluon interactions inside the nucleon or during the fragmentation process. A representative quark--gluon correlation function is the Qiu--Sterman function~[\cite{Qiu:1991pp, Qiu:1991wg}],
\begin{equation}
T_F(x,x) = \int \frac{d\lambda\,d\mu}{4\pi}\, e^{ix\lambda}\,
\langle P,S |\,
\bar\psi(0)\, \gamma^+ g_s F^{+\alpha}(\mu n)\, \psi(\lambda n)\, 
|\, P,S \rangle ,
\end{equation}
which plays a central role in the description of single-spin asymmetries. Such correlators provide direct information on color interactions between the active quark and the spectator system and have no analog within the naive parton model.

Experimentally, higher twists contribute to numerous azimuthal and spin-dependent asymmetries measured in SIDIS. For an unpolarized target, the $p_{h T}$-integrated cross section may be written schematically as~[\cite{Bacchetta:2006tn}]
\begin{equation}
\frac{d\sigma}
     {dx_B\,dy\,dz_h\,d\phi_h}
= F_{UU,T}
+ \varepsilon F_{UU,L} 
+ \sqrt{2\varepsilon(1+\varepsilon)} \cos\phi_h F_{UU}^{\cos\phi_h}
+ \varepsilon \cos2\phi_h F_{UU}^{\cos2\phi_h}
+ \cdots ,
\label{eq:sig_sidis}
\end{equation}
where $\phi_h$ is the azimuthal angle of the detected hadron and $\varepsilon$ is the virtual-photon polarization parameter. The first (second) subscript on the SIDIS structure functions denotes the beam (target) polarizations, with $U$ indicating an unpolarized beam or target and $L$ and $T$ corresponding to longitudinal and transverse polarization, while the superscripts specify the azimuthal modulation associated with the corresponding structure function. In Eq.~(\ref{eq:sig_sidis}) the $\cos\phi_h$ modulation contains both kinematic contributions associated with intrinsic transverse momentum (the Cahn effect) and genuine twist-3 quark--gluon correlations, while the $\cos2\phi_h$ term receives leading-twist contributions through the Boer--Mulders mechanism together with higher-order corrections.

Although the collinear factorization formalism provides a systematic framework for incorporating higher-twist contributions in $p_{h T}$-integrated SIDIS, there have been no phenomenological extractions of higher-twist distribution or fragmentation functions analogous to those performed for inclusive DIS. Such an extraction is naturally more challenging, since the SIDIS cross section depends simultaneously on PDFs and fragmentation functions, and requires precision data across a range of $x_B$, $z_h$ and $Q^2$ values in order to unambiguously identify the higher-twist effects. This is further complicated by the presence of hadron-mass effects, threshold corrections at large $z_h$, and possible violations of collinear factorization at moderate $Q^2$.

Experimental studies nevertheless provide some insight into the importance of higher-twist effects. Measurements of charged-pion electroproduction in Jefferson Lab Hall~C over the range $Q^2\approx2$--$4~{\rm GeV}^2$ demonstrated that the leading-order partonic description of semi-inclusive pion production remains surprisingly successful even in the nucleon resonance region~[\cite{Navasardyan:2006gv, Mkrtchyan:2007sr, Asaturyan:2011mq}]. These studies found that appropriately constructed pion production ratios exhibit quark--hadron duality and are broadly consistent with expectations from leading-twist collinear factorization. At the same time, the analyses emphasize that larger corrections are expected at lower $Q^2$, larger $x_B$, and in the limit $z_h\to1$, where threshold effects, hadron-mass corrections, and multiparton dynamics become increasingly important. 
The JAM Collaboration is currently the only group to incorporate $p_{h T}$-integrated SIDIS data into a global QCD analysis~[\cite{Anderson:2024evk, Cocuzza:2026zoy}], including charged-hadron multiplicities from the HERMES and COMPASS experiments, and is presently extending the analysis to include recent high-precision Jefferson Lab measurements of identified-hadron SIDIS cross sections, which cover the large-$x_B$, moderate-$Q^2$ region where power corrections are expected to be significantly larger than at HERMES and COMPASS kinematics.

A significant recent advance in the phenomenology of higher twists is the first global QCD analysis of genuine twist-three parton distributions by [\cite{Portela:2026wwn}]. In contrast to previous analyses that determine the twist-three structure function $g_2$ or its moment $d_2$ from polarized DIS data alone, their analysis simultaneously incorporates measurements of $g_2$, the $d_2$ moment, and the Sivers and worm-gear asymmetries in SIDIS within a unified collinear twist-three framework. An important motivation is that a single observable cannot uniquely determine the underlying twist-three quark--gluon correlation functions, which depend on two independent partonic momentum fractions. By combining inclusive and semi-inclusive measurements with the complete twist-three QCD evolution, the analysis is able to extract the underlying genuine twist-three distributions and demonstrates that a common set of universal correlation functions provides a consistent description of all available data. This represents the first phenomenological confirmation of the universality of twist-three factorization across both inclusive and semi-inclusive lepton scattering.

For polarized DIS, the analysis provides a detailed determination of the twist-three contribution to the structure function $g_2$ and the $d_2$ moment. At presently accessible energies, the twist-two Wandzura--Wilczek contribution is found to dominate the proton $g_2$ structure function, while the twist-three component becomes increasingly important for the neutron and at small $x_B$. At higher energies, however, the twist-three contribution is predicted to dominate over the Wandzura--Wilczek term across a much wider kinematic region, making future polarized measurements at the EIC especially sensitive to genuine quark--gluon correlations. The extracted values of $d_2$ are consistent with existing experimental measurements and lattice QCD calculations, and imply average transverse color Lorentz forces of approximately equal magnitude and opposite sign for $u$- and $d$-quarks. This work also illustrates the evolution of higher-twist phenomenology from the extraction of local OPE matrix elements in polarized DIS to global determinations of the underlying multiparton correlation functions that simultaneously describe inclusive
and semi-inclusive DIS.

\subsection{Higher-Twist TMD Distributions}
\label{sec:HT_TMD}

The situation is markedly different for observables that depend on the transverse momentum of the detected hadron. In this case, higher-twist effects appear directly through the $p_{hT}$ dependence of the SIDIS cross section and through a variety of azimuthal and spin asymmetries. Many of these observables either receive their first nonvanishing contribution at twist three or contain sizeable twist-three contributions that can be separated experimentally through their characteristic angular dependence. Consequently, the most direct experimental evidence for higher-twist dynamics in SIDIS comes from $p_{hT}$-differential measurements, where TMD factorization provides a natural framework for describing the underlying quark--gluon correlations. In this subsection we discuss higher-twist effects in $p_{hT}$-differential SIDIS and the associated twist-three TMD distributions.

Spin asymmetries provide particularly sensitive probes of higher-twist dynamics. Longitudinal beam-spin asymmetries receive contributions from twist-3 TMDs through interference between longitudinal and transverse virtual-photon amplitudes, while transverse target-spin asymmetries probe quark--gluon correlations associated with the Sivers and Collins mechanisms. Although these observables can often be described within either collinear twist-3 or TMD factorization, the two approaches are closely related in the kinematic region where both are applicable.

\begin{figure}[t]
\centering
\includegraphics[width=0.6\textwidth]{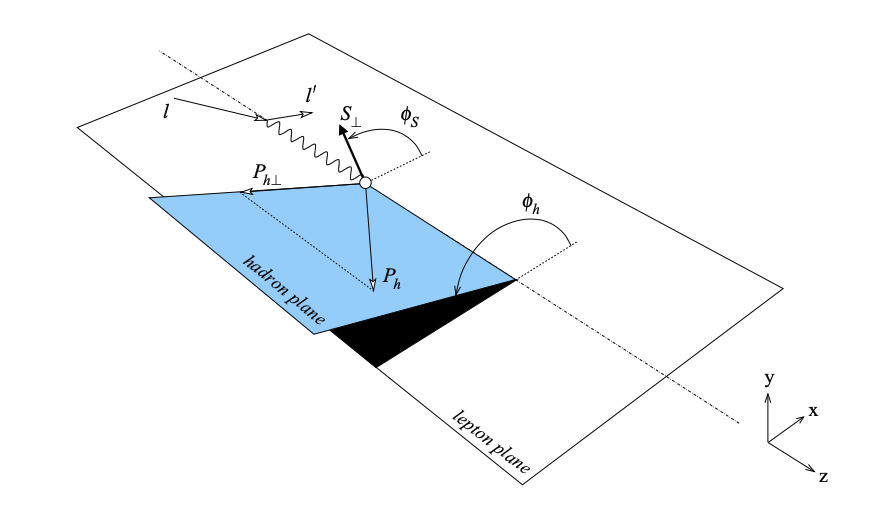}
\caption{Definition of azimuthal angles for the SIDIS process in the target rest frame according to the Trento convention, with $P_{h\perp}$ ($\equiv p_{hT}$) and $S_\perp$ the components of $p_h$ and $S$ transverse to the photon momentum. From [\cite{Bacchetta:2004jz}].}
\label{fig:sidis_geo}
\end{figure}

Figure~\ref{fig:sidis_geo} illustrates the geometry of the SIDIS process according to the Trento convention~[\cite{Bacchetta:2004jz}], defining the azimuthal angles of the detected hadron and target-spin vector with respect to the lepton scattering plane. At leading power only a limited number of angular modulations are allowed, while additional modulations appear at subleading order through twist-three quark--gluon correlations and kinematic effects suppressed by powers of $1/Q$.
For hadron transverse momenta satisfying $p_{hT}\ll Q$, the SIDIS cross section admits a TMD factorization in which the transverse momentum of the detected hadron arises from the intrinsic transverse momentum of the struck parton together with the transverse momentum generated during the fragmentation process. At leading power, the differential cross section may be written schematically as~[\cite{Collins:2011zzd, Bacchetta:2006tn, Metz:2016swz}].
\begin{align}
\frac{d\sigma}{dx_B\,dy\,dz_h\,d^2 p_{hT}}
=& \sigma_0 \sum_i e_i^2
\int d^2k_T\, d^2p_T\,
\delta^{(2)}\!\left( z_h{\bm k}_T+{\bm p}_T-{\bm p}_{hT} \right)
f_i(x_B,k_T^2;Q^2)\, D_i^h(z_h,p_T^2;Q^2)
+ {\cal O}\!\left(\frac{p_{hT}^2}{Q^2}\right),
\label{eq:TMDfactorization}
\end{align}
where $\sigma_0$ is the leading-order leptonic prefactor, ${\bm k}_T$ and ${\bm p}_T$ denote the transverse momenta of the incoming and outgoing partons, respectively, and $f_i(x_B,k_T^2;Q^2)$ and $D_i^h(z_h,p_T^2;Q^2)$ are the TMD parton distribution and fragmentation functions, respectively. Beyond leading order Eq.~(\ref{eq:TMDfactorization}) must be generalized to include the hard-scattering coefficient, soft factor, and TMD evolution.

The TMD formalism naturally organizes the SIDIS cross section according
to its twist expansion,
\begin{equation}
d\sigma 
= d\sigma^{(\tau=2)} 
+ \frac{M}{Q}\, d\sigma^{(\tau=3)}
+ {\cal O}\!\left(\frac{M^2}{Q^2}\right),
\label{eq:TMDtwistexpansion}
\end{equation}
where $d\sigma^{(\tau)}$ denotes the contribution from operators of twist $\tau$. The leading term involves the familiar twist-two TMDs (including the unpolarized, helicity, and transversity distributions), and the time-reversal-odd Sivers and Boer--Mulders functions. The first subleading contribution, suppressed by one power of $M/Q$, contains twist-three quark--quark and quark--gluon--quark correlation functions together with twist-three TMD fragmentation functions.

Experimentally, higher-twist TMDs can be accessed through azimuthal modulations of the SIDIS cross section. For an unpolarized target and longitudinally polarized lepton beam, the beam-spin asymmetry is defined
as
\begin{equation}
A_{LU}^{\sin\phi_h}
= \frac{F_{LU}^{\sin\phi_h}}{F_{UU,T} + \varepsilon F_{UU,L}},
\label{eq:ALU}
\end{equation}
where $F_{UU,T}$ and $F_{UU,L}$ are the ($p_{hT}$ dependent) unpolarized transverse and longitudinal SIDIS structure functions, respectively, and $F_{LU}^{\sin\phi_h}$ is the beam-spin structure function. Since $F_{LU}^{\sin\phi_h}$ first contributes at order $1/Q$, a nonzero beam-spin asymmetry provides direct evidence for subleading-power dynamics beyond the leading-twist approximation.

\begin{figure}[t]
\centering
\includegraphics[width=0.39\textwidth]{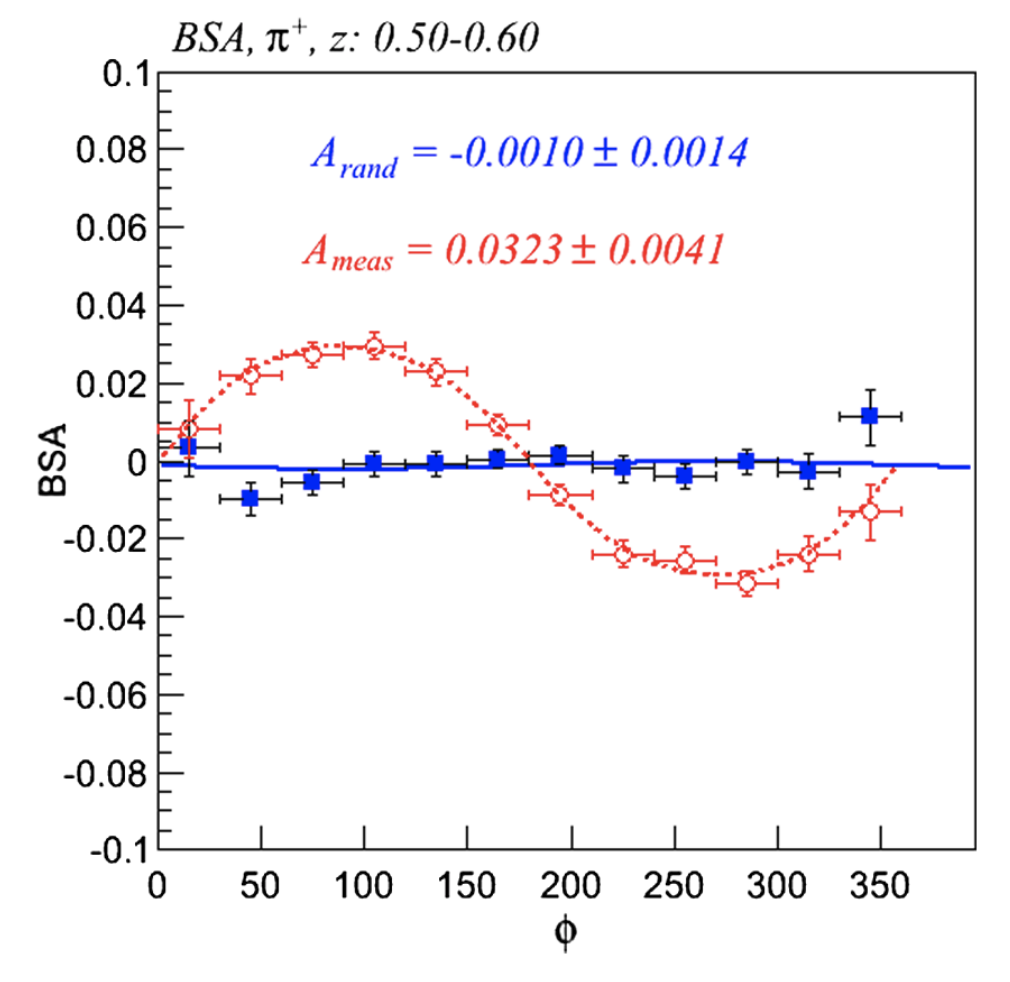}
\caption{Beam-spin asymmetry (BSA) $A_{LU}^{\sin\phi_h}$ versus $\phi_h$ ($=\phi$ in the figure) in semi-inclusive DIS from CLAS, for one bin in $z_h$ ($=z$ in the figure) using measured helicities (red circles), compared with the result obtained using random helicities (blue squares), which is expected to be zero, to estimate the systematic uncertainty. From~[\cite{CLAS:2014dmz}].}
\label{fig:ALUsidis}
\end{figure}

Figure~\ref{fig:ALUsidis} shows the beam-spin asymmetry $A_{LU}^{\sin\phi_h}$ measured by the CLAS Collaboration at Jefferson Lab~[\cite{CLAS:2014dmz}]. The asymmetry is clearly nonzero over the measured kinematic range, reaching magnitudes of several percent in the valence region. This provides one of the clearest experimental demonstrations of higher-twist effects in SIDIS, illustrating the presence of nontrivial quark--gluon correlations in the nucleon and in the fragmentation process~[\cite{Bacchetta:2006tn, Metz:2016swz}]. 
Comparable measurements have been performed by the HERMES and COMPASS Collaborations, confirming the existence of sizeable subleading-power azimuthal asymmetries over a broad kinematic range~[\cite{HERMES:2009lmz, COMPASS:2014kcy}]. Although no global extraction of the individual twist-three TMD distributions has yet been performed, these measurements provide important constraints on phenomenological models and demonstrate that higher-twist effects play an essential role in describing the spin-dependent structure of SIDIS.

In recent years considerable progress has also been made in establishing the connection between the TMD and collinear twist-three formalisms, where the transverse moments of certain TMDs can be related to the ETQS quark--gluon correlation functions~[\cite{Efremov:1981sh, Efremov:1984ip, Qiu:1991pp, Qiu:1991wg}]. This correspondence provides a unified description of single-spin asymmetries in the intermediate kinematic region $\Lambda_{\rm QCD}\ll p_{hT}\ll Q$, where both approaches are applicable~[\cite{Ji:2006ub, Koike:2007dg, Kang:2011hk}].\\

\subsection{Higher Twists in Exclusive Processes}
\label{sec:HT_exclusive}

Exclusive reactions provide a complementary probe of higher-twist dynamics by combining the information contained in ordinary PDFs and elastic form factors into generalized parton distributions (GPDs)~[\cite{Mueller:1994ses, Ji:1996ek, Radyushkin:1996nd}]. In contrast to inclusive DIS, where only the longitudinal momentum fraction is measured, GPDs depend additionally on the longitudinal momentum transfer (skewness) $\xi$ and the invariant momentum transfer $t$, thereby providing access to the spatial and momentum structure of partons inside hadrons. Higher-twist effects in exclusive processes arise through subleading hard-scattering amplitudes, twist-3 GPDs, higher-twist meson distribution amplitudes, and quark--gluon correlation functions. Comprehensive reviews of GPDs and deeply virtual exclusive processes can be found in~[\cite{Diehl:2003ny, Belitsky:2005qn, Kumericki:2016ehc}].

The theoretical description of deeply-virtual Compton scattering (DVCS) is based on the factorization of the Compton amplitude into perturbatively calculable coefficient functions and nonperturbative GPDs~[\cite{Ji:1998xh, Collins:1998be}]. The Compton form factors (CFFs), which parameterize the DVCS amplitude, are obtained by convoluting the perturbative coefficient functions with the corresponding GPDs. For the unpolarized GPD $H^q$, for example, one has
\begin{equation}
{\cal H}(\xi,t,Q^2)
= \sum_q e_q^2 \int_{-1}^{1}dx\, C(x,\xi,Q^2)\, H^q(x,\xi,t),
\label{eq:CFF}
\end{equation}
with analogous expressions for the CFFs of the other GPDs~[\cite{Belitsky:2005qn, Diehl:2003ny}]. Power corrections to the DVCS amplitude may be organized according to their twist,
\begin{equation}
{\cal T}
= {\cal T}^{(\tau=2)} 
+ \frac{1}{Q}\, {\cal T}^{(\tau=3)}
+ \frac{1}{Q^2}\, {\cal T}^{(\tau=4)} 
+ \cdots ,
\label{eq:DVCS_twist}
\end{equation}
where ${\cal T}^{(\tau)}$ denotes the contribution from operators of twist $\tau$. Unlike inclusive DIS, where the first dynamical corrections appear at order $1/Q^2$, exclusive reactions generally receive twist-3 contributions at order $1/Q$. These corrections originate from quark transverse momentum, quark--gluon correlations, and higher Fock-state components of the hadronic wave function~[\cite{Belitsky:2000vx, Kivel:2000fg}].

Phenomenologically, the most relevant observables in DVCS are beam-spin, beam-charge, and target-spin asymmetries, which arise from the interference between the DVCS and Bethe--Heitler (BH) amplitudes~[\cite{Belitsky:2001ns, Belitsky:2005qn, Kumericki:2016ehc}], and measurements of these asymmetries provide direct access to the real and imaginary parts of the CFFs. Higher-twist effects enter as subleading contributions to the interference term and can be isolated through the characteristic azimuthal dependence of the cross section.

The dependence of the cross section on the azimuthal angle $\phi$ between the lepton and hadron scattering planes can be expanded in a Fourier series,
\begin{equation}
\frac{d\sigma}{d\phi}
= c_0 
+ \sum_{n=1}^{\infty} 
  \left[ c_n\cos(n\phi) + s_n\sin(n\phi) \right],
\label{eq:DVCSharmonics}
\end{equation}
where the Fourier coefficients $c_n$ and $s_n$ depend on the beam and target polarizations and on the kinematic variables $x_B$, $Q^2$, and $t$. The harmonic decomposition reflects the helicity structure of the virtual Compton amplitude. At leading twist, the cross section is dominated by the $\cos\phi$ and $\sin\phi$ harmonics generated by the interference of the leading-twist handbag amplitude with the BH process. Twist-three amplitudes contribute to the $\cos2\phi$ and $\sin2\phi$ harmonics arising from longitudinal--transverse helicity interference, while higher harmonics are suppressed by additional powers of $1/Q$. Measurements of the azimuthal dependence therefore provide one of the most direct experimental probes of higher-twist dynamics in exclusive processes.

As an example, the lepton beam-spin asymmetry is approximately proportional to
\begin{equation}
A_{LU}
\propto
\frac{{\rm Im}\,{\cal H}}
{|{\cal T}_{\rm BH}|^2 + |{\cal T}_{\rm DVCS}|^2 + {\cal I}},
\label{eq:ALU_DVCS}
\end{equation}
where ${\cal H}$ is the dominant twist-two CFF, ${\cal T}_{\rm BH}$ and ${\cal T}_{\rm DVCS}$ denote the Bethe--Heitler and DVCS amplitudes, respectively, and ${\cal I}$ is their interference term. At leading power, the asymmetry is dominated by the $\sin\phi$ harmonic arising from the interference of the BH and leading-twist DVCS amplitudes. Twist-three CFFs and other higher-twist contributions generate subleading corrections to the dominant $\sin\phi$ modulation and contribute to higher Fourier harmonics, making precision measurements of the azimuthal dependence an important probe of higher-twist dynamics.

\begin{figure}[h]
\centering
\includegraphics[width=0.9\textwidth]{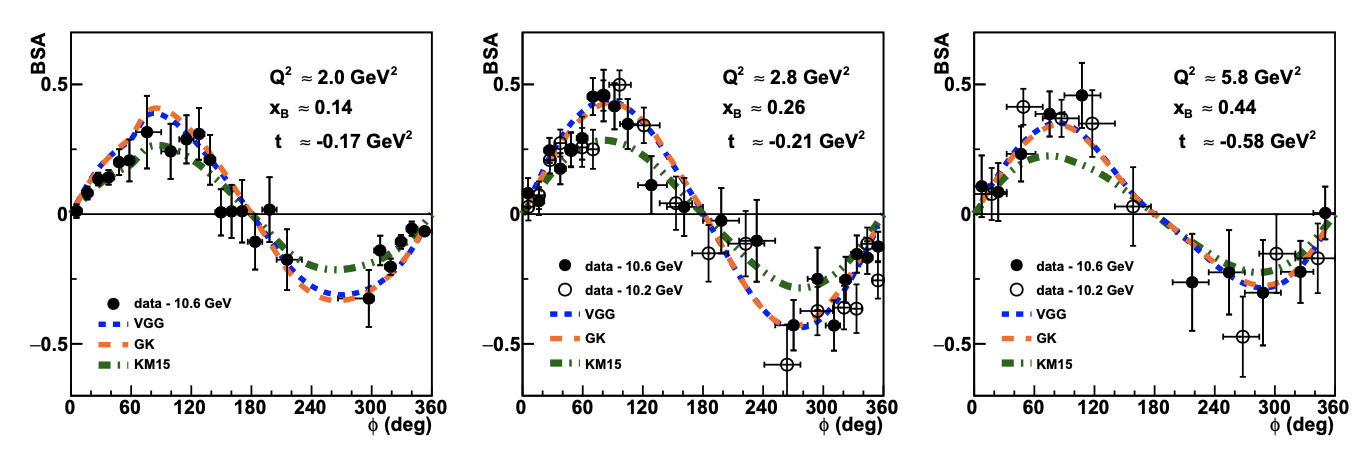}
\caption{Beam-spin asymmetry $A_{LU}$ for deeply virtual Compton scattering measured by the CLAS12 Collaboration as a function of the azimuthal angle $\phi$ in representative kinematic bins and compared with several GPD calculations. The observed $\sin\phi$ modulation is consistent with the leading-twist handbag mechanism, while precision measurements of the azimuthal dependence provide sensitivity to subleading higher-twist contributions. Adapted from~[\cite{CLAS:2022syx}].}
\label{fig:ALUdvcs}
\end{figure}

Figure~\ref{fig:ALUdvcs} shows the beam-spin asymmetry measured by the CLAS12 Collaboration~[\cite{CLAS:2022syx}] as a function of the azimuthal angle $\phi$ for representative kinematic bins. The observed asymmetries exhibit the characteristic $\sin\phi$ dependence expected from the leading-twist handbag mechanism and are well described by modern GPD calculations over the explored kinematic range. Although the measured asymmetries are dominated by the leading Fourier harmonic, the increasing precision of Jefferson Lab data will enable progressively more stringent tests of subleading $1/Q$ corrections through departures from the leading $\sin\phi$ behavior and the extraction of higher Fourier harmonics.

Exclusive meson production provides another important laboratory for studying higher twists. While longitudinal vector-meson production factorizes at leading twist, transverse vector mesons and pseudoscalar meson production receive significant twist-3 contributions through both higher-twist meson distribution amplitudes and quark--gluon correlations~[\cite{Collins:1996fb, Anikin:2009bf}]. These effects are essential for describing many polarization observables measured at HERA, COMPASS, and Jefferson Lab.
Over the past two decades, measurements by HERMES, H1, ZEUS, COMPASS, and Jefferson Lab have established deeply-virtual exclusive reactions as precision tools for investigating the three-dimensional structure of hadrons. Although most phenomenological analyses are based on leading-twist factorization, the steadily improving experimental precision has made the inclusion of higher-twist effects increasingly important for extracting reliable GPDs and CFFs~[\cite{Kumericki:2016ehc}]. The future Electron--Ion Collider will considerably extend the kinematic coverage and provide stringent tests of the onset of the leading-twist regime and the role of twist-3 dynamics in exclusive reactions. \\

\section{Summary and Outlook}
\label{sec:summary}

Higher-twist contributions constitute an essential component of the QCD description of lepton--hadron scattering over a broad range of kinematics. While suppressed by powers of $1/Q^2$ relative to the leading-twist contribution, higher twists encode fundamentally new information on nonperturbative quark--gluon interactions and multiparton correlations. Rather than representing merely corrections to the asymptotic Bjorken scaling behavior, higher twists provide direct insight into the emergence of color confinement and the transition between perturbative and nonperturbative QCD.
In this review we have summarized the theoretical foundations, phenomenology, and current status of higher-twist studies in DIS and related hard scattering processes. Beginning with the OPE, we discussed how the moments of structure functions may be organized into a systematic expansion in inverse powers of $Q^2$, with leading-twist operators describing incoherent scattering from individual partons and higher-twist operators characterizing coherent multiparton interactions. The separation of TMCs from genuine dynamical higher twists remains essential for obtaining physically meaningful determinations of nonperturbative matrix elements from experimental data.

Phenomenological analyses of inclusive DIS have demonstrated that higher-twist effects are relatively modest over much of the kinematic range explored by modern experiments, but become increasingly important at large values of $x_B$ and moderate $Q^2$, where perturbative and nonperturbative dynamics overlap. Over the past several decades, high-precision measurements from facilities around the world have enabled increasingly quantitative extractions of effective higher-twist contributions to nucleon structure functions, highlighting the sensitivity of extracted higher-twist coefficients to higher-order perturbative QCD corrections, TMCs, threshold resummation, heavy-quark effects, and nuclear corrections.

An important theme has been the close connection between higher twists and quark--hadron duality. The remarkable observation that resonance-region averages reproduce scaling structure functions over a surprisingly wide kinematic range suggests that coherent multiparton effects remain relatively small even in regions where individual resonances dominate the cross section. Understanding this behavior continues to provide valuable insight into the onset of perturbative QCD and the interplay between partonic and hadronic degrees of freedom.

Beyond inclusive DIS, higher twists play a central role in many other hard scattering processes. In SIDIS they contribute through twist-3 and twist-4 TMD PDFs and fragmentation functions, generating characteristic azimuthal asymmetries and spin-dependent observables that probe quark transverse motion and quark--gluon correlations. In TMD factorization they provide an increasingly rich description of spin-orbit correlations inside hadrons, while in exclusive processes such as DVCS and exclusive meson production they are essential for describing transverse photon amplitudes and restoring electromagnetic gauge invariance at finite values of $Q^2$.

Future theoretical challenges will involve improved understanding of operator mixing, higher-order coefficient functions, and factorization beyond leading power, among other areas of research. Similarly, phenomenological analyses will require increasingly sophisticated treatments to consistently incorporate higher-order perturbative corrections, target mass effects, threshold resummation, and electroweak radiative corrections. Future global analyses will likely combine these ingredients with flexible parametrizations, Bayesian inference techniques, and machine learning methods to provide more reliable determinations of higher-twist contributions and their uncertainties.

An important development will be the rapidly growing connection between phenomenological analyses and first-principles lattice QCD calculations. Traditional lattice calculations of local operator
matrix elements are now being complemented by quasi-PDF and pseudo-PDF
approaches that provide direct information on the $x$ dependence of
parton distributions. As lattice calculations continue to improve in
precision and are increasingly incorporated into global QCD analyses,
they offer the prospect of constraining higher-twist matrix elements
directly from QCD rather than relying exclusively on phenomenological
fits to experimental data.

On the experimental front, the completion of the Jefferson Lab 12~GeV program will provide unprecedented measurements of nucleon structure functions, polarized observables, semi-inclusive processes, and exclusive reactions throughout the valence region, where higher-twist effects are expected to be largest. Further ahead, the Electron--Ion Collider will extend these investigations over a vastly broader kinematic range, enabling precision studies of the transition from the perturbative to the nonperturbative regime, while providing stringent tests of QCD factorization beyond leading twist. The combination of high-luminosity measurements with polarized beams and a wide range of nuclear targets will open entirely new opportunities for studying multiparton correlations and the dynamics of color confinement.

Taken together, advances in perturbative QCD, lattice QCD simulations, global QCD analyses, and precision experiments are rapidly transforming higher-twist physics from a qualitative description of power corrections into a quantitative tool for investigating the nonperturbative structure of hadrons. Continued progress in these areas promises not only increasingly precise determinations of higher-twist matrix elements but also a deeper understanding of the emergence of hadronic structure from the fundamental quark and gluon degrees of freedom of QCD.

\begin{ack}[Acknowledgments]

I thank Jianwei Qiu for helpful comments and suggestions.
This work was supported by the U.S. Department of Energy, Office of Science, Office of Nuclear Physics under Contract No. 89243126CSC000213.

\end{ack}

\bibliographystyle{Harvard}
\bibliography{reference}

\end{document}